\documentclass[aps,superscriptaddress,prd,
onecolumn,
floatfix,
nofootinbib,
longbibliography,
amsmath,amssymb,amsfonts]{revtex4-2}
\usepackage{amsmath,amssymb}
\usepackage{amsthm}
\usepackage{mathrsfs}
\usepackage[pdftex]{graphicx}

\usepackage{bm}
\usepackage{subfigure}
\usepackage{float}
\usepackage{physics,cancel}

\usepackage{comment}
\usepackage{braket}
\theoremstyle{plain}
\newtheorem{thm}{Theorem}
\newtheorem{cor}{Corollary}

\newcommand\nn{\nonumber \\}
\newcommand{\pa}{\partial}

\let\Re\relax
\DeclareMathOperator{\Re}{Re}
\let\Im\relax
\DeclareMathOperator{\Im}{Im}
\usepackage[colorlinks,linkcolor=blue,anchorcolor=violet,citecolor=red]{hyperref}

\newcommand{\anno}[1]{\textcolor{red}{#1}}

\newcommand{\annom}[1]{\textcolor{magenta}{#1}}
\begin{document}
\title{Entanglement Harvesting from Quantum Field: Insights via the Partner Formula}
\author{Yuki Osawa}
\email{osawa.yuki.e8@s.mail.nagoya-u.ac.jp}
\affiliation{ Department of Physics, Nagoya University, Nagoya 464-8602, Japan}
\author{Yasusada Nambu}
\email{nambu.yasusada.e5@f.mail.nagoya-u.ac.jp}
\affiliation{ Department of Physics, Nagoya University, Nagoya 464-8602, Japan}
\author{Riku Yoshimoto}
\email{yoshimoto.riku.d1@s.mail.nagoya-u.ac.jp}
\affiliation{ Department of Physics, Nagoya University, Nagoya 464-8602, Japan}
\date{\today}
\begin{abstract}
We examine the conditions under which entanglement can be extracted from a quantum field using two localized modes A and B (detector modes). Building upon Simon's entanglement criterion \cite{simon2000peres} for bipartite Gaussian states, we reformulate the condition in terms of commutators between the canonical operators of detector mode B and the partner mode P of detector mode A. This reformulation allows us to derive a no-go theorem for entanglement harvesting in certain settings, including uniformly accelerating detectors. Drawing upon the known correspondence between the Unruh effect and Hawking radiation, our result implies the absence of quantum correlations between ``real particles" emitted as Hawking radiation.
\end{abstract}

\maketitle
\tableofcontents

\section{Introduction}
 The quantum field theory in black hole spacetimes predicts the emission of thermal Hawking radiation from black holes, the temperature of which is given by the surface gravity at their event horizons \cite{Hawking:1975vcx,hawking1974black}. 
 The key to this thermal radiation is the existence of the Killing horizon and the analyticity of the quantum field across the Killing horizon \cite{wald1994quantum,birrell1984quantum}.
Thus, the emergence of thermal radiation is not unique to black holes; we encounter similar phenomena for the uniformly accelerating observer in flat spacetime \cite{unruh1976notes,candelas1977vacuum,takagi1986vacuum,unruh1984happens,landulfo2019classical,ievlev2024electron,gallock2025acceleration}, flat spacetime with an accelerating boundary \cite{fulling1976radiation,davies1977radiation,carlitz1987reflections}, an expanding universe \cite{gibbons1977cosmological}, or even in analog gravity systems \cite{unruh1981experimental,unruh2005universality,busch2014quantum,nambu2023entanglement,osawa2023particle,steinhauer2016observation,hotta2022expanding,PhysRevD.107.085002}.

 From the perspective of quantum information, the thermal property of the quantum field implies the existence of a purification partner that is entangled with radiation and purifies it completely. 
In spacetimes possessing a Killing horizon, the purification partner of thermal radiation is located inside the horizon, leading to the presence of non-local correlations across the horizon.
To illustrating this phenomenon, Reznik proposed a protocol, that extracts such non-local correlations from quantum field vacuum \cite{reznik2003entanglement,reznik2005violating,summers1987maximal}.
He demonstrated that two uniformly accelerating Unruh-DeWitt (UDW) detectors \cite{unruh1976notes,dewitt1979,birrell1984quantum} in flat Minkowski spacetime can become entangled, and concluded that the origin of entanglement is the non-local correlation of the quantum field vacuum.
Today, this protocol --known as entanglement harvesting-- serves as a powerful tool in the field of relativistic quantum information. It has been extensively studied in various settings, focusing on the possibility and amount of entanglement that can be extracted by this protocol, as these quantities reflect the structure of non-local correlations of the quantum field \cite{benatti2010entangling,nambu2013entanglement,simidzija2018harvesting,henderson2018harvesting,cong2019entanglement,simidzija2017nonperturbative,gallock2021harvesting,henderson2022entanglement,de2023entanglement}. 

\begin{figure}[H]
\centering
\includegraphics[width=0.5\linewidth]{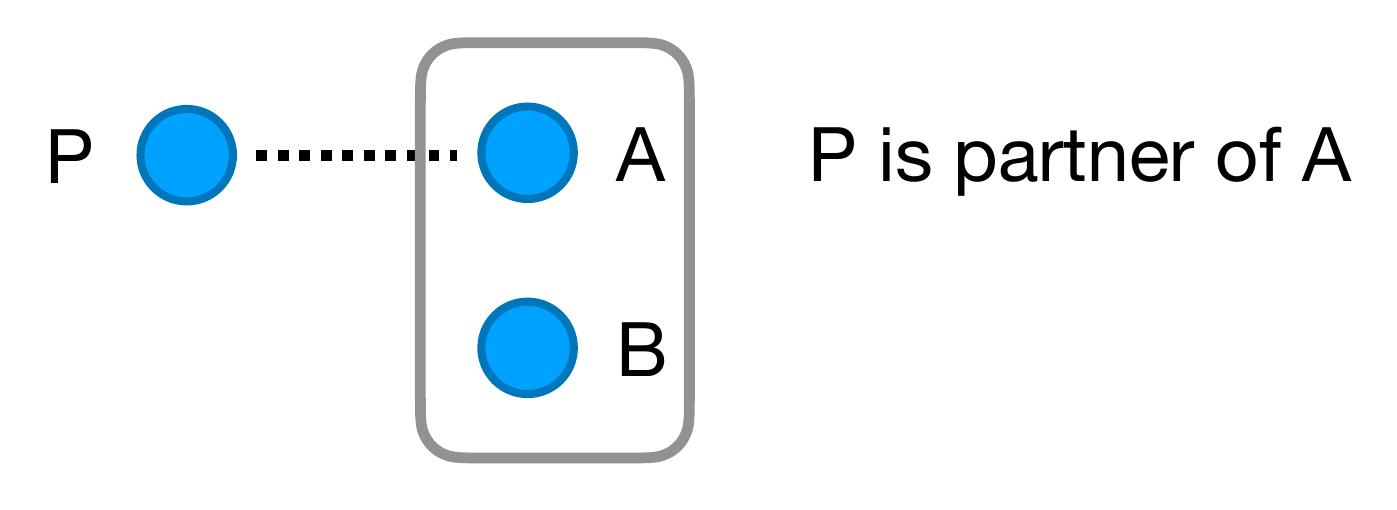}
\caption{Setup of our problem. We introduce local modes A and B (detector modes) from the quantum field and investigate entanglement  of the bipartite state AB. P denotes the partner mode of A that purifies mode A.}
\label{fig:setup}
\end{figure}

The partner formula \cite{hotta2015partner, trevison2019spatially, hackl2019minimal,yamaguchi2020superadditivity,nambu2023entanglement} may serve as an effective tool for investigating entanglement harvesting. 
This formula identifies the partner mode that completely purifies a given mode. 
In the harvesting procedure, we require a pair of interactions corresponding to the UDW detectors.
Using the partner formula, we can construct the complementary interaction that enables the extraction of entanglement with maximum efficiency for a given interaction \cite{trevison2018pure}.
The procedures are as follows (Fig.~\ref{fig:setup}).
When the details of an interaction between the detector and the quantum field are specified, we can identify the mode A detected by that particle detector, often referred to as the detector mode \cite{trevison2019spatially,tomitsuka2020partner,osawa2024final} of the interaction. 
Then, by applying the partner formula to the detector mode A, we obtain its partner mode P and can construct the second interaction, whose detector mode coincides with this partner mode.
Since the detector mode A and its partner mode P are maximally entangled by their construction, these interactions allow us to harvest the maximum amount of entanglement.
However, in practical scenarios of entanglement harvesting, it is inappropriate to classify the interactions as arbitrary, since the interactions between quantum fields and detectors are typically subject to constraints.
Therefore, it may not always be possible to choose interactions such that one detector mode is the partner mode of the other.
Nevertheless, we expect that the partner formula will provide a useful criterion for determining the feasibility of entanglement harvesting, as accessing information about the partner mode is essential for extracting entanglement.
We intuitively believe that the profile functions \cite{trevison2019spatially,tomitsuka2020partner,osawa2024final} of the partner mode A of the detector mode A correspond to the spatial location of the partner, and we can extract entanglement by using detector modes A and B when the profile functions of B and P overlap. However, a definitive entanglement criterion based on the partner formula has not yet been established.

In this paper, we discuss the criterion for successful entanglement harvesting from the perspective of the partner formula. 
We begin by analyzing Simon's entanglement criterion (the entanglement criterion that appears in the Lemma of Simon's paper \cite{simon2000peres}) for detector modes A and B. By reformulating it in terms of commutators between detector mode B and the partner mode P of detector A, we derive a no-go theorem that imposes fundamental limitations on the extraction of entanglement using detector modes A and B.
Based on the profile representation of detector modes, we demonstrate that the reformulated Simon's criterion partly supports our intuitive understanding of the relationship between the partner formula and the feasibility of entanglement harvesting: the overlap of the profile functions of the detector modes P and B is necessary for extracting entanglement using detector modes A and B. 
Interestingly, however, detector modes A and B can be separable even when the profile functions of P and B overlap, which may seem counterintuitive.
This is because both the overlap condition and Simon's criterion are necessary conditions for entanglement, with the overlap condition being weaker than Simon's criterion.

A central example illustrating our no-go theorem is entanglement harvesting in uniformly accelerated systems, specifically using detector modes composed of Rindler modes in the right wedge (Region I) of Minkowski spacetime. This case is not only a concrete application of the reformulated Simon's criterion but also highlights the broader implications for field-theoretic phenomena such as the Unruh effect and Hawking radiation.
By examining the profile functions of the detector mode, we observe that for a single Rindler mode, its profile and the profile of the partner appear on the opposite side of the horizon \cite{carlitz1987reflections,hotta2015partner}. In contrast, when the detector mode A is defined as a linear combination of the Rindler modes, the profile of the partner mode P leaks across the horizon due to the nonlinearity of the partner relation \cite{hotta2015partner,osawa2024final}.
Intuitively, it may seem possible to extract entanglement by using two detector modes A and B consisting solely of the Rindler modes in Region I, as the profile functions of the partner mode P and that of B overlap. However, reformulated Simon's criterion reveals that the detector modes A and B are separable when they consist only of positive frequency Rindler modes. 
Notably, the entanglement structure of the Rindler modes discussed here is closely related to the vacuum fluctuation scenario \cite{wilczek1993quantum, carlitz1987reflections,hotta2015partner,tomitsuka2020partner,wald2019particle,osawa2024final} of black hole evaporation, in which Hawking radiation is purified without an energy cost.
In light of this connection, and by analogy between Hawking radiation and the Unruh effect, our finding implies that no quantum correlation exists between ``real particles” emitted as Hawking radiation during the early stage of black hole evaporation, since these real particles emitted as Hawking radiation correspond to the positive frequency Rindler modes in this analogy. 

The organization of this paper is as follows:
In Sec.~\ref{sec-introdtecmode}, we briefly introduce the concept of the UDW detector and the detector mode based on a simple toy model.
In Sec.~\ref{sec:parametrization}, we parametrize Gaussian local modes using squeezing and rotation parameters and characterize the covariance matrix in terms of these parameters. In Sec.~\ref{sec:condent}, we reformulate Simon's entanglement criterion using the partner formula.
In Sec.~\ref{sec:profilerep}, we introduce the profile representation of local modes and present the entanglement criterion in this representation.
In Sec.~\ref{sec:appltoUnruh}, we apply the criterion to the quantum field in Rindler spacetime.
Sec.~\ref{sec:conclusion} is devoted to the conclusion.
Throughout this paper, we use natural units $c=\hbar=k_B=1$.

\section{Quantum Measurement on Quantum Fields and the Detector Mode}\label{sec-introdtecmode}
The process of extracting quantum information from a quantum system is called measurement. In quantum theory, even with repeated applications of the same measurement on identically prepared systems, it is impossible to obtain complete information about the quantum state. Furthermore, the information accessible through measurement is determined only after the specific measurement scheme has been specified.
The von Neumann measurement, or indirect measurement, is one framework for describing quantum measurement processes. 
In this framework, an ancillary quantum system called a probe, which interacts with the target quantum system, is introduced. After the interaction, the transition probability of the probe system is measured, and from this result, the initial quantum state of the target system can be indirectly inferred.

This is also the case for quantum fields. In this context, the detector model serves as an example of a von Neumann measurement applied to a quantum field, where the UDW detector \cite{unruh1976notes,dewitt1979,birrell1984quantum} acts as the probe system. 
Once the detector model is specified, the accessible information about the quantum field state is represented by the detector mode associated with that detector model.
Indeed, the detector mode encapsulates the detector’s action on the quantum field, allowing the detector model to be defined by specifying the detector mode rather than by detailing the UDW detector itself (e.g., the characteristics of the probe system or the nature of its interaction with the field).
In this section, we briefly describe the detector model for  quantum field measurement and introduce the concept of the detector mode using a simple toy model.

As a demonstration, we consider a massless scalar field coupled to the UDW detector, which consists of two energy levels. The Hamiltonian of the system is given by
\begin{align}
    \hat{H}&=\hat{H}_\phi+\hat{H}_D+\hat{H}_{int},\nn
    &\equiv \frac{1}{2}\int d^3{x}\ \left(\hat{\pi}^2+(\nabla{\hat{\phi}})^2\right)+\Omega\,{\hat{\sigma}_z}+\lambda\chi(\tau){\hat{\sigma}_x}\,\hat{\phi}(\tau,\xi_0),
\end{align}
where $\hat{H}_\phi$ and $\hat{H}_D$ represent the free parts of the field and the detector, respectively, and $\hat{H}_{int}$ describes the interaction between them. Here, $ \hat\sigma_{x,y,z}$ are Pauli matrices and $\tau$ denotes the proper time of the detector. The terms $\hat{H}_D$ and $\hat{H}_{int}$ include the parameters of the UDW detector, such as the energy gap $\Omega$, the coupling constant $\lambda$ between the systems, and the switching function $\chi(\tau)$ governing the interaction, while $\xi_0$ specifies the detector’s position in the comoving frame.

For simplicity, let us assume a flat Minkowski spacetime, with the UDW detector fixed at $x=0$ in the Minkowski coordinates.
As $\tau=t$ in this situation, the time evolution operator $\hat{U}$ after the interaction is given by
\begin{align}
    \hat{U}&=T\exp\left(-i\int^\infty_{-\infty}\dd{t'} \hat{H}^I_{int}(t')\right),\nn
    &\equiv T\exp\left(-i\lambda\int^\infty_{-\infty}\dd{t'} \chi(t')\left(e^{i\Omega t'}\hat{\sigma}_++e^{-i\Omega t'}\hat{\sigma}_-\right)\hat{\phi}(t',0)\right),\nn
    &\approx1-i\lambda\int^\infty_{-\infty}\dd{t'} \chi(t')\left(e^{i\Omega t'}\hat{\sigma}_++e^{-i\Omega t'}\hat{\sigma}_-\right)\hat{\phi}(t',0),
\end{align}
where $\hat{H}^I_{int}$ is the interaction Hamiltonian in the interaction picture, and $\hat\sigma_{\pm}=\hat\sigma_x\pm i\hat\sigma_y$ are the ladder operators. In the last line, the terms of order $\mathcal{O}(\lambda^2)$ are neglected under the assumption that the interaction between the probe and the quantum field is weak.
By adopting the mode expansion of the field operator, the time evolution operator can be further simplified:
\begin{align}
    \hat{U}&=1-i\lambda\int_0^\infty\dd\omega\left\{\left(\tilde{\chi}(\Omega-\omega)\hat{a}_\omega+\tilde{\chi}(\Omega+\omega)\hat{a}_\omega^\dag\right)\hat{\sigma}_++h.c.\right\},\nn
    &\equiv1-i\lambda\mathcal{N}\left(\hat{A}_D\hat{\sigma}_++h.c.\right),
\end{align}
where $\tilde{\chi}(\omega)$ is the Fourier transform of the switching function $\chi(t)$ with respect to the mode function defined as
\begin{align}
    \tilde{\chi}(\omega)=\int_{-\infty}^\infty\dd{t'}\chi(t')\frac{e^{i\omega t'}}{\sqrt{4\pi\omega}},
\end{align}
and $\mathcal{N}$ is a normalization constant chosen such that the new annihilation operator $\hat{A}_D$ satisfies the commutation relation $[\hat{A}_D,\hat{A}_D^\dag]=1$ .

We note that, at this stage, the detector can access only the subspace of the Fock space spanned by the creation and annihilation operators $\hat{A}_D$ and $\hat{A}^\dag_D$. Hence, the annihilation operator $\hat{A}_D$ represents the degrees of freedom accessible to the detector under consideration. This can be seen more clearly if we assume that the initial state of the detector is its ground state. When this detector state is coupled to some unknown field state $|\Psi\rangle_\phi$, the transition probability of the detector is given by
\begin{align}
    \text{Prob}\left(|g\rangle_D\to|e\rangle_D\right)=\lambda^2\mathcal{N}^2\langle\Psi|\hat{A}_D^\dag\hat{A}_D|\Psi\rangle_\phi.
\end{align}
Therefore, the detector measures the number of particles associated with $\hat{A}_D$, but it cannot extract any other information about the field state.

As the relation between annihilation operators $\left\{\hat{a}_\omega\right\}_\omega$ and $\hat{A}_D$ is defined through a Bogoliubov transformation, $\hat{A}_D$ defines the  mode function $\varphi_D$ of the quantum field as
\begin{align}
    \varphi_D(t,x)=\left[\hat{\phi}(t,x),\hat{A}_D^\dag\right].
\end{align}
The mode specified by $\hat{A}_D$ is referred to as the detector mode. 
The detector mode encapsulates all the information accessible to the detector under consideration and allows us to discuss quantum operations on the field without specifying the detailed structure of the UDW detector.
The same applies to discussions of entanglement harvesting: the possibility of harvesting can be analyzed by examining the entanglement structure of the detector modes.
For this reason, we focus solely on the detector modes in the following sections.

\section{Parametrization of Modes and Entanglement}\label{sec:parametrization}

As we shall see in Sec.~\ref{sec:profilerep}, the detector modes of the quantum field are expressed as linear combinations of the annihilation and creation operators of the quantum field, as the field operators can be expanded in terms of these operators.
Hence, the detector modes of the quantum field can be analyzed within the framework of Gaussian quantum information(quantum information for harmonic oscillators). In this section, we will construct a general framework applicable to various oscillator modes before discussing the quantum field.

Let us consider two independent oscillator modes $\hat{a}_A$ and $\hat{a}_B$ embedded in $n$-mode Gaussian modes specified by the annihilation operators $\{\hat{b}_1,\cdots,\hat{b}_n\}$:
\begin{align}
    \label{eq:annirep}
    &\hat{a}_A:=\sum_{i=1}^n \left\{\alpha_i^A\hat{b}_i+\beta_i^A(\hat{b}_i)^\dag\right\},\quad
    \hat{a}_B:=\sum_{i=1}^n \left\{\alpha_i^B\hat{b}_i+\beta_i^B(\hat{b}_i)^\dag\right\}, \\
&\sum_i(|\alpha_i^A|^2-|\beta_i^A|^2)=1,\quad\sum_i(|\alpha_i^B|^2-|\beta_i^B|^2)=1.
\end{align}
By performing a basis transformation (see Appendix~\ref{ap-derivation of parametrization} for the detailed calculation), $\hat{a}_A$ and $\hat{a}_B$ 
can be rewritten as:
\begin{align}
    \label{eq:defmodeA}
	\hat{a}_A&=\cosh r\,\hat{a}_\parallel+\sinh r\,\hat{a}^\dag_\perp,\\
    \label{eq:defmodeB}
	\hat{{a}}_B&=\cosh r_1\left(\cos\theta_1\,\hat{a}_\parallel+\sin\theta_1\cos\xi_1\,\hat{a}_\perp+\sin\theta_1\sin\xi_1\,\hat{a}_0\right)\nn
	&+\sinh r_1\left(\cos\theta_2\,\hat{a}_\parallel^\dag+\sin\theta_2\cos\xi_2\,\hat{a}_\perp^\dag+\sin\theta_2\sin\xi_2\cos\chi\,\hat{a}_0^\dag+\sin\theta_2\sin\xi_2\sin\chi\,\hat{a}_1^\dag\right),
\end{align}
where $\hat{a}_\parallel$ and $\hat{a}_\perp$ denote independent modes that decompose the given mode A to the superposition of  two modes, and $\hat{a}_0$ and $\hat{a}_1$ are bases of the oscillator modes which are orthogonal to $\hat{a}_\parallel$ and $\hat{a}_\perp$.
To ensure the independence of the $\hat{a}_A$ and $\hat{{a}}_B$, we require
\begin{align}
	\label{eq:indep-1}
	0&=[\hat{a}_A,\hat{{a}}_B]=\cosh r\sinh r_1\cos\theta_2-\sinh r\cosh r_1\sin\theta_1\cos\xi_1,\\
	\label{eq:indep-2}
	0&=[\hat{a}_A^\dag,\hat{a}_B]=-\cosh r\cosh r_1\cos\theta_1+\sinh r\sinh r_1\sin\theta_2\cos\xi_2.
\end{align}
Using these relations, we can determine the parameters $\xi_1$ and $\theta_1$ for given $r, r_1,\theta_2,\xi_2$. 

The covariances of the canonical operators $(\hat{q},\hat{p})$ are known to serve as indicators of the quantumness of the system \cite{simon2000peres, simon1994quantum}. 
In particular, for Gaussian systems, they are related to the Wigner function, which has a one-to-one correspondence with the density matrix of the quantum state, and the uncertainty relation can be reformulated in terms of these covariances.
Furthermore, the covariances are also effective for evaluating quantum entanglement between two Gaussian modes. In the following, we evaluate the entanglement between modes A and B by employing the covariance of the canonical operators.

For this purpose, we first introduce canonical operators $\hat{q}_A,\hat{p}_A,\hat{q}_B,\hat{p}_B$ corresponding to modes A and B:
\begin{equation}
\hat q_A=\frac{\hat a_A+\hat a_A{}^\dag}{\sqrt{2}},
\quad\hat p_A=\frac{\hat a_A-\hat a_A{}^\dag}{i\sqrt{2}},
\quad\hat q_B=\frac{\hat a_B+\hat a_B{}^\dag}{\sqrt{2}},
\quad\hat p_B=\frac{\hat a_B-\hat a_B{}^\dag}{i\sqrt{2}}.
\end{equation}
We can verify that the canonical commutation relation $[\hat{q}_i,\hat{p}_j]=i\delta_{ij}$ is automatically satisfied.
The covariances of the canonical operators can be summarized in matrix form as follows, which is referred to as the covariance matrix:
\begin{align}
	V^{(AB)}\equiv\left(\begin{array}{cc}
	V_{AA} &V_{AB}\\
	V_{AB}^T&V_{BB}
	\end{array}\right)
    =\left(\begin{array}{cccc}
	 \langle\{\hat{q}_A,\hat{q}_A\}\rangle & \langle\{\hat{q}_A,\hat{p}_A\}\rangle & \langle\{\hat{q}_A,\hat{{q}}_B\}\rangle & \langle\{\hat{q}_A,\hat{{p}}_B\}\rangle \\ 
	 \langle\{\hat{p}_A,\hat{{q}}_A\}\rangle &  \langle\{\hat{p}_A,\hat{{p}}_A\}\rangle & \langle\{\hat{p}_A,\hat{{q}}_B\}\rangle & \langle\{\hat{{p}}_A,\hat{{p}}_B\}\rangle \\ 
	 \langle\{\hat{q}_B,\hat{{q}}_A\}\rangle & \langle\{\hat{q}_B,\hat{{p}}_A\}\rangle & \langle\{\hat{{q}}_B,\hat{{q}}_B\}\rangle & \langle\{\hat{q}_B,\hat{p}_B\}\rangle \\
	  \langle\{\hat{{p}}_B,\hat{q}_A\}\rangle & \langle\{\hat{{p}}_B,\hat{{p}}_A\}\rangle & \langle\{\hat{{p}}_B,\hat{q}_B\}\rangle & \langle\{\hat{{p}}_B,\hat{{p}}_B\}\rangle
	  \end{array}\right),
	  \label{eq:covstandard}
\end{align}
where the expectation value is evaluated for the quantum state under consideration(e.g., 
$\langle\{\hat{q}_A,\hat{q}_A\}\rangle
:=\Tr[\{\hat{q}_A,\hat{q}_A\}\hat{\rho}])$.

In the following, we consider the vacuum state $\ket{0000}_{\parallel\perp01}$ with $\hat{a}_\parallel\ket{0000}_{\parallel\perp01}=\hat{a}_\perp\ket{0000}_{\parallel\perp01}=\hat{a}_0\ket{0000}_{\parallel\perp01}=\hat{a}_1\ket{0000}_{\parallel\perp01}=0 $ and entanglement extraction from this vacuum state with two detector modes A and B. The non-vanishing elements of the covariance matrix for this quantum state are given as follows: 
\begin{align}
	\langle\{\hat{q}_A,\hat{{q}}_A\}\rangle&=\langle\{\hat{p}_A,\hat{{p}}_A\}\rangle=\cosh 2r,\\
	 \langle\{\hat{q}_A,\hat{{q}}_B\}\rangle&=2\sinh r_1(\sinh r\sin\theta_2\cos\xi_2+\cosh r\cos\theta_2),\\
	 \langle\{\hat{p}_A,\hat{{p}}_B\}\rangle&=2\sinh r_1(\sinh r\sin\theta_2\cos\xi_2-\cosh r\cos\theta_2),\\
	\langle\{\hat{{q}}_B,\hat{{q}}_B\}\rangle&=(\cosh r_1\cos\theta_1+\sinh r_1\cos\theta_2)^2+(\cosh r_1\sin\theta_1\cos\xi_1+\sinh r_1\sin\theta_2\cos\xi_2)^2\nn
	&\vspace{2cm}+(\cosh r_1\sin\theta_1\sin\xi_1+\sinh r_1\sin\theta_2\sin\xi_2\cos\chi)^2
	+\sinh^2 r_1\sin^2\theta_2\sin^2\xi_2\sin^2\chi,\\
	\langle\{\hat{{p}}_B,\hat{{p}}_B\}\rangle&=(\cosh r_1\cos\theta_1-\sinh r_1\cos\theta_2)^2+(\cosh r_1\sin\theta_1\cos\xi_1-\sinh r_1\sin\theta_2\cos\xi_2)^2\nn
	&\vspace{2cm}+(\cosh r_1\sin\theta_1\sin\xi_1-\sinh r_1\sin\theta_2\sin\xi_2\cos\chi)^2+\sinh^2 r_1\sin^2\theta_2\sin^2\xi_2\sin^2\chi,
\end{align}

The essence of the entanglement structure for two-mode Gaussian state can be captured by two specific sets of parameters, with which the covariance matrix reduces to the standard form of a bipartite Gaussian modes system. 
The covariance matrix takes the conventional standard form with the choice of the parameters $\theta_1=\pi/2,\theta_2=0,\cos\xi_1=\tanh r_1/\tanh r$ or $\theta_1=0,\theta_2=\pi/2,\cos\xi_1=1/(\tanh r_1\tanh r)$:
\begin{equation}
    V^{(AB)}_\text{standard}=\begin{bmatrix}
    a&0&c_1&0\\ 0&a&0&c_2\\
    c_1&0&b&0\\ 0&c_2&0&b
    \end{bmatrix},
\end{equation}
For the former choice of parameters, the matrix elements are given by 
 $a=\cosh 2r$, $b=\cosh 2r_1$, and $c_1=-c_2=2\cosh r\sinh r_1$.
This state reduces to a pure two-mode state  (two-mode squeezed state), which is the maximally entangled state, especially when $r=r_1,\xi_1=0$. This can be seen more clearly by writing down the explicit form of the bipartite mode for this set of parameters:
\begin{align}
\hat{a}_A&=\cosh r\,\hat{a}_\parallel+\sinh r\,\hat{a}^\dag_\perp, \label{eq:modeA}\\
\hat{{a}}_B&=\cosh r_1\left(\cos\xi_1\,\hat{a}_\perp+\sin\xi_1\,\hat a_0\right)
	+\sinh r_1\,\hat{a}_\parallel^\dag. \label{eq:modeB}
\end{align}

Under the former choice of the parameters, the covariance matrix reduces to the standard form with $c_1c_2<0$, whereas it reduces to the standard form with $c_1c_2\ge0$ by the latter choice of parameters(i.e., $\theta_1=0,\theta_2=\pi/2, \cos\xi_2=1/(\tanh r\tanh r_1)$). In this case, $c_1=c_2=2\cosh r\cosh r_1$, and  mode B is expressed as
\begin{equation}
\hat a_B=\cosh r_1\,\hat a_\parallel+\sinh r_1\cos\xi_2\,\hat a_\perp^\dag+\sinh r_1(\sin\xi_2\cos\chi\,\hat a_0^\dag+\sin\xi_2\sin\chi\,\hat a_1^\dag).
\end{equation}
Note that this state does not reduce to a pure two-mode state for any choice of parameters $r_1$ and $\xi_2$, as $\hat{a}_B$ does not contain terms with $\hat{a}_{\parallel}^\dag$ and $\hat{a}_\perp$. Considering the two examples above and recalling that $\det V_{AB}=c_1c_2$, we may be tempted to conclude that the sign of $\det V_{AB}$ indicates the presence of two mode squeezing between $\hat{a}_A$ and $\hat{a}_B$, which appears to be necessary for entanglement between A and B. Indeed, this expectation corresponds precisely to Simon's criterion \cite{simon2000peres}.

We can quantify the entanglement between two modes using  the PPT criterion \cite{peres1996separability,horodecki1997separability,simon2000peres},
which provides the necessary and sufficient condition for the entanglement of {the bipartite Gaussian state}. 
In this criterion, we evaluate the symplectic eigenvalues of the partially transposed covariance matrix to test the PPT criterion; the existence of negative eigenvalues provides the necessary and sufficient condition for entanglement in bipartite Gaussian states. In some situations, it is sufficient to apply a weaker entanglement criterion, which is easier to check.
Simon showed \cite{simon2000peres} that the following  is a necessary condition for the entanglement of the bipartite state AB:
\begin{align}
	\det V_{AB}<0.
    \label{eq:Simon}
\end{align}

\section{Condition for Entanglement and Its Relation with the Partner Formula}\label{sec:condent}
Based on the partner formula  \cite{hotta2015partner}, the mode that  purifies the mode $\hat{a}_A$ in Eq.~\eqref{eq:defmodeA} is given by
\begin{align}
	\hat{a}_P=\cosh r \,\hat{a}_\perp+\sinh r\,\hat{a}_\parallel^\dag.
	\label{eq:partnerformula}
\end{align}
Actually, using the canonical variables $\hat\xi_j=(\hat q_A,\hat p_A,\hat q_P, \hat p_P)$, the covariance matrix for the bipartite state AP becomes
\begin{equation}
(V^{(AP)})_{jk}=\expval{\{\hat\xi_j,\hat\xi_k\}}
=\begin{pmatrix}
\cosh 2r & 0 & \sinh 2r & 0 \\
0 & \cosh 2r & 0 & -\sinh 2r \\
\sinh 2r & 0 & \cosh 2r & 0 \\
0 & -\sinh 2r & 0 & \cosh 2r
\end{pmatrix},
\end{equation}
which is a pure two-mode squeezed state.
In the intuitive sense, the possibility of entanglement harvesting with detector modes $\hat{a}_A$ and $\hat{a}_B$ is related to ``how the detector mode $\hat{a}_B$ and the partner mode $\hat{a}_P$ resemble each other''. 
Hotta \textit{et al.} used the commutation relation between the creation and annihilation operators of two modes B and P as an inner product of two modes \cite{hotta2015partner} to show the independence of these modes. 
Following their strategy, we consider two types of commutators between the modes $\hat{a}_B$ and $\hat{a}_P$:
\begin{align}
	[\hat{{a}}_B,\hat{a}_P^\dag]&=-\sinh r\sinh r_1\cos\theta_2+\cosh r\cosh r_1\sin \theta_1\cos\xi_1, \label{eq:aBP1}\\
	[\hat{a}_B,\hat{a}_P]&=\sinh r\cosh r_1\cos\theta_1-\cosh r\sinh r_1\sin\theta_2\cos\xi_2. \label{eq:aBP2}
\end{align}
Using the independence conditions \eqref{eq:indep-1} and \eqref{eq:indep-2} of $\hat{a}_A$ and $\hat{a}_B$, we obtain
\begin{align}
	[\hat{a}_B,\hat{a}_P^\dag]&=\frac{\cos\theta_2}{\sinh r}\sinh r_1,\\
	[\hat{a}_B,\hat{a}_P]&=-\frac{\sin\theta_2\cos\xi_2}{\cosh r}\sinh r_1.
\end{align}
However, these are not invariants under the local symplectic transformations of each mode.
In fact, under the following local transformation of the detector mode B
\begin{align}
	\hat{{a}}'_B=e^{i\phi_1}\cosh {r}'\,\hat{a}_B+e^{i\phi_2}\sinh {r}'\,\hat{a}_B^\dag,
\end{align}
although the local transformation should not affect the correlation between $\hat{a}_A$ and $\hat{a}_B$, the values on the right-hand side of Eqs.~\eqref{eq:aBP1} and \eqref{eq:aBP2} are entirely altered.  

To find quantities that are invariant under local symplectic transformations, we focus on the covariance matrix $V=V^{(AB)}$ and its symplectic transformation under the canonical transformation of modes:
\begin{align}
	{V}'=SVS^T, 
\end{align}  
where $S$ is the symplectic matrix satisfying
\begin{align}
	SJS^T=J,
	\  \ J=\left(\begin{array}{cc}
	0 &1\\
	-1&0
	\end{array}\right)\oplus\left(\begin{array}{cc}
	0 &1\\
	-1&0
	\end{array}\right).
\end{align}
The local symplectic transformation is represented by
\begin{align}
	S=S_A\oplus S_B,
\end{align}
where ${S}_A$ and $S_B$ are the symplectic matrices for  single modes A and B, respectively.
Applying this transformation, the covariance matrix is transformed as
\begin{align}
	{V}'=\begin{pmatrix}
	S_A\,V_{AA}\,S_A^T & S_A\,V_{AB}\,S_B^T\\
	S_B\,V^T_{AB}\,S_A^T & S_B\,V_{BB}\,S_B^T
	\end{pmatrix}.
\end{align}
From the symplectic condition $\det S=1$,  we have four symplectic invariants $\det V, \det V_{AA}, \det V_{AB},\det V_{BB}$, that are invariant under the local symplectic transformation.
With our parametrization,
 the determinants of covariance sub-matrices are explicitly given as
\begin{align}
    \label{eq:detVAA}
    \det V_{AA}&=\cosh^2 2r,\quad \det V_{BB}=\cosh^2 2r_1\\
    \label{eq:detVAB}
    \det V_{AB}&=-4(\cosh^2r\cos^2\theta_2-\sinh^2r\sin^2\theta_2\cos^2\xi_2)\sinh^2 r_1,
\end{align}
where the independence conditions \eqref{eq:indep-1} and \eqref{eq:indep-2} are used to eliminate $\theta_1$ and $\xi_1$ in $\mathrm{det}\, V_{AB}$.

On the other hand, we notice that the quantity $
	-\left|[\hat{a}_B,\hat{a}_P^\dag]\right|^2+\left|\left[\hat{a}_B,\hat{a}_P\right]\right|^2 $
is also invariant under the local mode transformation;
 Recalling equations Eqs.~(\ref{eq:detVAA}) and (\ref{eq:detVAB}), we can directly check that this discriminant can be connected to the symplectic invariants as 
\begin{align}
	D:=-\left|[\hat{a}_B,\hat{a}_P^\dag]\right|^2+\left|[\hat{a}_B,\hat{a}_P]\right|^2=\frac{\det V_{AB}}{\det V_{AA}-1}.
    \label{eq:D}
\end{align}
Noting that $\det V_{AA}\ge1$, Simon's necessary condition for  entanglement can be rewritten as
\begin{align}
	D<0.
    \label{eq:Dneg}
\end{align}
Therefore, as one application of the concept of entanglement partner and the partner formula, it is possible to test whether two detector modes satisfy the necessary condition for entanglement. 
We can also show that the criterion using the discriminant $D$ is applicable even when the detector mode A contains single mode squeezing (see Appendix~\ref{sec:app-generalize} for the proof in this case):
\begin{align}
    \hat{a}_A:=\cosh r\,\hat{a}_\parallel+\sinh r\cos\theta\,\hat{a}_\parallel^\dag+\sinh r\sin\theta\,\hat{a}_\perp^\dag.
\end{align}
Hence, we can use the criterion Eq.~\eqref{eq:Dneg} even if the definitions of the modes are different from Eq.~(\ref{eq:defmodeA}) or Eq.~(\ref{eq:defmodeB}), since any covariance matrix of the bipartite Gaussian state can be transformed to Eq.~(\ref{eq:covstandard}) by the local mode transformation. 
\section{Profile Function Representations of the Local Modes and the Entanglement Criterion}\label{sec:profilerep}

In the paper by Trevison \textit{et al.}  \cite{trevison2019spatially},
the detector mode and its partner mode of the quantum field are represented as smeared field operators. 
Indeed, their representation of the detector mode by using smearing function (profile representation) reduces to the  representation in terms of annihilation operators (annihilation operator representation) such as Eq.~\eqref{eq:annirep}. Hence, the method discussed in the previous sections can also be applied to detector modes defined through the profile representation.
In this section, we briefly review their approach and show how Simon’s criterion from the previous section can be formulated in terms of the profile functions that define the smeared field operators.

Here, we start from a pair of the canonical operators $\{\hat{Q}_A,\hat{P}_A\}$ defined through smeared field operators in $(1+1)$-dimensional spacetimes:
\begin{align}
    \label{eq:profilerepq}
	\hat{Q}_A&:=\int dx \left\{V_A(x)\,\hat{\phi}(t,x)+W_A(x)\,\hat{\Pi}(t,x)\right\},\\
    \label{eq:profilerepp}
	\hat{P}_A&:=\int dx \left\{X_A(x)\,\hat{\phi}(t,x)+Y_A(x)\,\hat{\Pi}(t,x)\right\},
\end{align}
where  $\hat{\Pi}=\pa_t\hat{\phi}$ is the conjugate momentum of the field operator $\hat{\phi}$, and
 $V_A(t,x)$, $W_A(t,x)$, $X_A(t,x)$, and $Y_A(t,x)$ are  real profile functions that satisfy the condition
\begin{align}
	\int dx \left\{V_A(x)Y_A(x)-W_A(x)X_A(x)\right\}=1,
\end{align}
which comes from the canonical commutation relation $[\hat{Q}_A,\hat{P}_A]=i$. This pair of canonical operators defines the measurements or other local operations on the quantum field; conversely, a given local operation determines the corresponding pair of canonical operators or profile functions. 

As in the case of harmonic oscillators, we define the annihilation operator $\hat{a}_A$ as 
\begin{align}
	\hat{a}_A:=\frac{\hat{Q}_A+i\hat{P}_A}{\sqrt{2}}.
\end{align}
Using the canonical commutation relation, we can confirm that the commutation relation for the creation and annihilation operators $[\hat{a}_A,\hat{a}_A^\dag]=1$ is satisfied.
By using the mode expansion of the field operator
\begin{equation}
	\hat{\phi}(t,x)=\int_0^\infty d\omega \,(\varphi_\omega(t,x)\hat{a}_\omega+\varphi^*_\omega(t,x)\hat{a}^\dag_\omega),\quad [\hat{a}_{\omega_1},\hat{a}^\dag_{\omega_2}]=\delta_{\omega_1,\omega_2},
    \label{eq:fmode}
\end{equation}
the  operator $\hat{a}_A$ can be expressed as the linear combination of the creation operators $\hat{a}_\omega$:
\begin{align}
    \label{eq:annirep2}
	\hat{a}_A&=\frac{1}{\sqrt{2}}\int dx\, d\omega\Bigl[\left(V_A(x)+i X_A(x)\right)\varphi_\omega(t,x)+\left(W_A(x)+i Y_A(x)\right)\pa_t\varphi_\omega(t,x)\Bigr]\hat{a}_\omega\nn
	&\quad +\frac{1}{\sqrt{2}}\int dx \,d\omega\Bigl[\left(V_A(x)+i X_A(x)\right)\varphi_\omega^*(t,x)+\left(W_A(x)+i Y_A(x)\right)\pa_t\varphi_\omega^*(t,x)\Bigr]\hat{a}_\omega^\dag.
\end{align}

Therefore, the local operations on the quantum field,  represented by a pair of canonical operators $\{\hat{Q}_A,\hat{P}_A\}$ or by the profile functions $\{V_A(t,x), W_A(t,x),X_A(t,x),Y_A(t,x)\}$, can be expressed in terms of the Bogoliubov transformation between the annihilation operators $\hat{a}_A$ and $\{\hat{a}_\omega\}$. 
Conversely, once the relation between the annihilation operators $\hat{a}_A$ and $\{\hat{a}_\omega\}$ is specified, the profile functions can be reconstructed by following the procedure described above in reverse (see Sec.~\ref{sec:V-B} and Appendix~\ref{sec:app-derprof} for details).
Indeed, $\hat{a}_A$ is represents the detector mode associated with the local operation, and Eqs.~(\ref{eq:profilerepq}) and (\ref{eq:profilerepp}) is simply the different representation of the this detector mode. The expression in terms of annihilation operators (Eq.~\eqref{eq:annirep} or Eq.~\eqref{eq:annirep2})is referred to as the annihilation operator representation, while the one using profile functions (Eqs.~(\ref{eq:profilerepq}) and (\ref{eq:profilerepp})) is referred to as the profile representation.

Now, let us consider entanglement harvesting from the quantum field using two detector modes A and B, defined by profile functions. The most straightforward way to specify these modes would be to define all the profile functions associated with modes A and B. However, we adopt a different strategy here:
the detector mode B is first configured using profile functions, as
\begin{align}
	\hat{Q}_B&:=\int d x \left\{V_B(x)\,\hat{\phi}(t,x)+W_B(x)\,\hat{\Pi}(t,x)\right\},\\
	\hat{P}_B&:=\int d x \left\{X_B(x)\,\hat{\phi}(t,x)+Y_B(x)\,\hat{\Pi}(t,x)\right\}.
\end{align}
When applying Simon's criterion, our interest lies in the partner mode of the detector mode A rather than in the mode A itself. Therefore, we specify the detector mode A indirectly through its partner mode P, instead of specifying it directly:
\begin{align}
	\hat{Q}_P&:=\int d x \left\{V_P(x)\,\hat{\phi}(t,x)+W_P(x)\,\hat{\Pi}(t,x)\right\},\\
	\hat{P}_P&:=\int d x \left\{X_P(x)\,\hat{\phi}(t,x)+Y_P(x)\,\hat{\Pi}(t,x)\right\}.
\end{align}
This approach is justified by the one-to-one correspondence between the detector mode A and its partner mode P, whose profile functions are related through the partner formula\cite{hotta2015partner,trevison2019spatially,tomitsuka2020partner,hackl2019minimal,osawa2024final}.
We then apply Simon's necessary condition for bipartite entanglement, Eq.~\eqref{eq:Simon}, by evaluating $D$ defined in Eq.~\eqref{eq:D}:
\begin{align}
	D:&=-\left|\left[\hat{a}_P,\hat{a}_B^\dag\right]\right|^2+\left|\left[\hat{a}_P,\hat{a}_B\right]\right|^2\nn
	&=2\left(-[\hat{Q}_P,\hat{Q}_B][\hat{P}_P,\hat{P}_B]^*+[\hat{P}_P,\hat{Q}_B][\hat{Q}_P,\hat{P}_B]^*\right),
\end{align}
where we used the fact that the commutator of canonical operators is purely imaginary to derive the last line.
Expressing the commutators in terms of the profile functions, we have:
\begin{align}
&[\hat Q_P,\hat Q_B]=i\int dx\,(V_PW_B-W_PV_B),\quad
[\hat P_P,\hat P_B]=i\int dx\,(X_PY_B-Y_PX_B),\\
&[\hat P_P,\hat Q_B]=i\int dx\,(X_PW_B-Y_PV_B),\quad
[\hat Q_P,\hat P_B]=i\int dx\,(V_PY_B-W_PX_B).
\end{align}

We observe that these commutators vanish
when the pairs of the profile functions $\{V_B,W_B,X_B,Y_B\}$ and\\ $\{V_P,W_P,X_P,Y_P\}$ have no overlap. In this case, $D=0$ and harvesting entanglement from a quantum field cannot be achieved through the detector modes A and B. Hence, an overlap between the profile functions of mode B and the profile functions of partner mode P of A, which correlates with the detector mode A, is necessary for extracting entanglement using the detector modes A and B.
Consideration in this section reinforces our intuitive understanding of the profile function representation of the local modes: the support of the detector mode's profile functions corresponds to the location of the detector mode, while the support of the partner mode's profile functions corresponds to the location of the partner.
\section{Application to the Unruh Effect and Hawking radiation}\label{sec:appltoUnruh}
In the previous subsection, we observed that the overlap between detector mode B and the partner P of another detector mode A provides one criterion for  entanglement harvesting, since the overlap of these profiles is necessary for the entanglement. However, overlap of the profile functions  generally does not provide a sufficient condition. We can see this in the example of the superposed Rindler modes. 
In this section, we only consider the left-moving mode of a chiral quantum field.

\subsection{Rindler mode, Its Partner and Unruh Effect}
\begin{figure}[ht]
	\begin{center}
        \includegraphics[width=7cm]{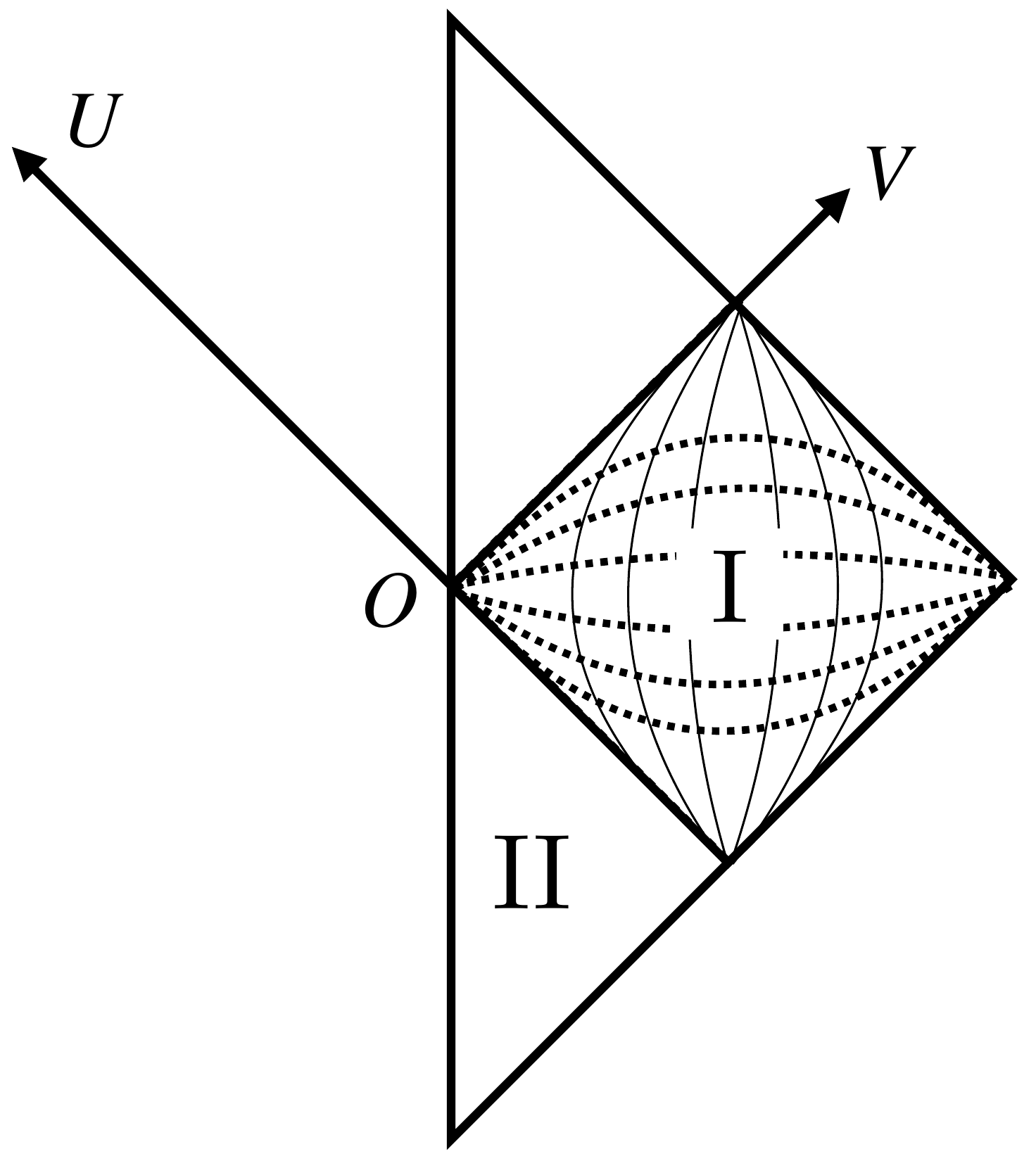}
        \caption{Penrose diagram depicting one-half of  the Minkowski spacetime. In region I, the solid lines denote surfaces of constant $\xi$, while the dashed lines indicate surfaces of constant $\eta$.}
        \label{fig:paraneg}
        \end{center}
\end{figure}
In this subsection, we briefly review the Unruh effect \cite{wald1994quantum,birrell1984quantum,osawa2024final}. 
We consider a massless scalar field in the $(1+1)$-dimensional Minkowski spacetime, satisfying the Klein-Gordon equation $\square\phi=0$. 
We begin with an examination of the structure of Minkowski spacetime before proceeding to analyze the scalar field. The metric is
\begin{align}
	ds^2=-dt^2+dx^2.
\end{align} 
For later use, we introduce null coordinates $U=t-x,\ V=t+x$.
The metric can be expressed as $ ds^2=-dUdV$.
Now, let us consider  new coordinates $(\eta,\xi)$ that are related to Minkowski coordinates $(t,x)$ by
\begin{align}
	t&=\frac{1}{a}e^{a\xi}\sinh a\eta,\quad
	x=\frac{1}{a}e^{a\xi}\cosh a\eta,\quad -\infty<\eta<+\infty,\quad -\infty<\xi<+\infty.
\end{align}
This new coordinate system is called the Rindler coordinates, which are the coordinates associated with a uniformly accelerating observer.
The metric with the Rindler coordinates is given by 
\begin{align}
	ds^2=-e^{2a\xi}(d\eta^2-d\xi^2).
\end{align}
The important property of the Rindler coordinates is that these coordinates cover only region I of Minkowski spacetime (see Fig. \ref{fig:paraneg}). Thus, the accelerating observer can only see part of the spacetime.
 We introduce null coordinates for the Rindler coordinates $u=\eta-\xi,~v=\eta+\xi$.
The metric with the null coordinates is expressed as 
\begin{align}
	ds^2=-e^{2a\xi}\,du\,dv.
\end{align}
The null coordinates of Minkowski spacetime and Rindler coordinates are related as
\begin{align}
	U&=-\frac{1}{a}\,e^{-au},\quad V=\frac{1}{a}\,e^{av}.
\end{align}

Using the null coordinates, the field equation can be written as $\pa_U\pa_V\phi(U,V)=0$
in the Minkowski coordinates and $\pa_u\pa_v\phi(u,v)=0$ in the Rindler coordinates.
In the Rindler coordinates, the left-moving mode function in region I is defined by 
\begin{align}
	\phi^{\text{I}}_\omega(V):&=\frac{1}{\sqrt{4\pi|\omega|}}\exp(-i\omega v(V))\,\theta(V) \notag \\
	&=\frac{1}{\sqrt{4\pi|\omega|}}\exp\left(-i\frac{\omega}{a}\log(aV)\right)\theta(V),\quad \omega>0,
\end{align}
 where we multiplied the step function $\theta(V)$
to ensure that the Rindler coordinates cover only region I and the mode function $\phi_\omega^\text{I}$ has support only in region I. It is also possible to introduce a mode function in region II  defined by
\begin{align}
	\phi^{\text{II}}_\omega(V):&=\frac{1}{\sqrt{4\pi|\omega|}}\exp(-i\omega v(-V))\theta(-V) \notag \\
	&=\frac{1}{\sqrt{4\pi|\omega|}}\exp\left(-i\frac{\omega}{a}\log(-aV)\right)\theta(-V),\quad \omega>0.
\end{align}
We notice that $\phi^{\text{II}}_{\omega}(V)$ has  a negative norm, and to obtain the positive norm modes, we need to take the complex conjugate of this mode. In fact, the positive norm mode $\phi^{\text{II}}_{-\omega}(V)=(\phi^{\text{II}}_{\omega}{}(V)){}^*$ is known as the Milne mode.

Although the mode functions $\phi^{\text{I}}_\omega$ and $\phi^{\text{II}}_\omega$ are non-analytic, the following functions defined by linear combinations of $\phi^\text{I}_\omega$ and $\phi^\text{II}_{\omega}$ are analytic in the entire Minkowski spacetime:
\begin{align}
\Phi_\omega(V)&:=\frac{e^{\pi\omega/2a}\,\phi^{\text{I}}_\omega(V)+e^{-\pi\omega/2a}\,(\phi^{\text{II}}_{-\omega}(V))^*}{\sqrt{2\sinh(\pi\omega/a)}}, \\
	\Psi_{-\omega}(V)&:=\frac{e^{-\pi\omega/2a}\,\left(\phi^\text{I}_\omega(V)\right)^*+e^{\pi\omega/2a}\,\phi^{\text{II}}_{-\omega}(V)}{\sqrt{2\sinh(\pi\omega/a)}}.
\end{align}
These functions are bounded everywhere in the lower-half complex $V$ plane and are orthogonal to each other. The modes defined by these functions are called Unruh modes.
The analytic property of the Unruh modes implies that these modes correspond to the positive norm Minkowski modes with respect to the Klein-Gordon inner product.
By solving these relations for $\phi^\text{I}_\omega$ and $\phi^\text{II}_\omega$, we obtain 
\begin{align}
		\label{eq:bogrind}
		\phi^\text{I}_\omega(V)&=\alpha_\omega\Phi_\omega(V)-\beta_\omega\left(\Psi_{-\omega}(V)\right)^*,\\
		\label{eq:bogrindpart}
		\phi^{\text{II}}_{-\omega}(V)&=-\beta_\omega\left(\Phi_\omega(V)\right)^*+\alpha_\omega\,\Psi_{-\omega}(V),
\end{align}
and the coefficients are given by
\begin{align}
	\alpha_\omega&:=\frac{e^{\pi\omega/2a}}{\sqrt{2\sinh (\pi\omega/a)}},\quad
		\beta_\omega:=\frac{e^{-\pi\omega/2a}}{\sqrt{2\sinh (\pi\omega/a)}},\quad\alpha_\omega^2-\beta_\omega^2=1.
\end{align}

We introduce annihilation operators by using the Klein-Gordon inner product\footnote{In this article, we adopt the definition of the KG inner product as $(f,g):=-i\int dV\{f\partial_Vg^*-g^*\partial_Vf\}$(KG inner product for null Cauchy surface $U=const$.).} between the field operator $\hat{\phi}$  and positive norm mode functions:
\begin{align}
	\hat{a}^\text{I}_\omega&=(\hat{\phi},\phi^\text{I}_\omega),
	\hspace{0.5cm}\hat{a}^\text{II}_\omega=(\hat{\phi},\phi^\text{II}_{-\omega}),\\
	\hat{a}^\Phi_\omega&=(\hat{\phi},\Phi_\omega),
	\hspace{0.5cm}\hat{a}^\Psi_\omega=(\hat{\phi},\Psi_{-\omega}).
\end{align}
Using the relations for the mode functions, the corresponding annihilation operators are related as
\begin{align}
		\label{eq:bogrindop}
		\hat{a}^\text{I}_\omega&=\alpha_\omega\,\hat{a}^\Phi_\omega+\beta_\omega\left(\hat{a}^\Psi_\omega\right)^\dag,\\
		\label{eq:bogrindpartop}
		\hat{a}^{\text{II}}_{\omega}&=\beta_\omega\left(\hat{a}_\omega^\Phi\right)^\dag+\alpha_\omega\,\hat{a}^\Psi_{\omega}.
\end{align}
If we take $r$ such that $\sinh r=\beta_\omega$, the first formula Eq.~(\ref{eq:bogrindop}) coincides with Eq.~(\ref{eq:defmodeA}) and the second formula Eq.~(\ref{eq:bogrindpartop}) coincides with its partner formula Eq. (\ref{eq:partnerformula}). Thus, the Milne mode $\hat a_\omega^\text{II}$ is the partner mode of the Rindler mode $\hat a_\omega^\text{I}$.

Indeed, $\hat{a}^\text{I}_\omega$ coincides with the detector mode measured by the uniformly accelerating observer. 
To see this, let us consider a UDW detector with two internal energy levels moving with a constant acceleration in Minkowski spacetime.
The interaction between the detector and the scalar field in the interaction picture is assumed to be
\begin{align}
	\hat{H}_{int}^{\text{I}}(\tau)={\lambda} \,(e^{i\Omega \tau}\,\hat{\sigma}_++e^{-i\Omega \tau}\,\hat{\sigma}_-)\,\hat{\phi}(\tau,\xi_0),
\end{align}
where $\lambda$ is a coupling constant between the scalar field and the detector, $\Omega$ is the energy gap of the detector, $\hat{\sigma}_\pm$ are ladder operators, and $\xi_0=\text{const.}$ is the trajectory (world line) of the detector in  the comoving coordinate system of the detector. 
It is important to note that the energy gap $\Omega$ should be defined with respect to the comoving frame of the UDW detector.
The time evolution operator is then expressed as
\begin{align}
	\hat{U}\simeq 1-i{\lambda}\int d{\tau'}(e^{i\Omega \tau'}\hat{\sigma}_++e^{-i\Omega \tau'}\hat{\sigma}_-)\,\hat{\phi}(\tau',\xi_0)
    =1-i{\lambda}\frac{e^{i\Omega\xi_0}}{\sqrt{4\pi\Omega}}\,\hat{\sigma}_+\,\hat{a}^\text{I}_\Omega-i{\lambda}\frac{e^{-i\Omega\xi_0}}{\sqrt{4\pi\Omega}}\,\hat{\sigma}_-\left(\hat{a}^\text{I}_\Omega\right)^\dag.
\end{align}
When evaluating the time integral, the mode expansion with respect to the detector's proper time must be chosen. 
In this case, the creation and annihilation operators of the Rindler mode appear, since the detector's proper time $\tau$ corresponds to the Rindler time $\eta$.

Since the mode functions $\Phi_\omega$ and $\Psi_{-\omega}$ only contain the positive frequency Minkowski modes, the vacuum state defined by $\hat{a}^\Phi_\omega$, $\hat{a}^\Psi_{\omega}$ and the Minkowski vacuum $|0\rangle_M$ are the same vacuum state.
Therefore, if we prepare the initial state of the quantum field as the Minkowski vacuum state and the initial state of the detector as the ground state (i.e. a state satisfying $\hat{\sigma}_-|0\rangle_D=0$), the excitation probability of the detector is 
\begin{align}
	\mathrm{Prob}\left(|0\rangle_D\to|1\rangle_D\right)\propto\frac{1}{\Omega}\,|\beta_\Omega|^2=\dfrac{1}{\Omega}\dfrac{1}{e^{2\pi\Omega/a}-1}.
\end{align}
The factor $1/(e^{2\pi\Omega/a}-1)$ represents the Planckian distribution with a temperature $T=a/2\pi$. The same thermal factor appears if we consider the transition probability of the detector immersed in the thermal bath with a temperature of $T$. Therefore, we cannot distinguish between the quantum field seen by the accelerating observer and the thermal quantum field using detectors at rest. This  is called the Unruh effect.

\subsection{The Superposition of Rindler Modes and Its Overlapped Partner}\label{sec:V-B}
In this subsection, we consider the detector whose mode is given by a superposition of the Rindler modes:
\begin{align}
	\hat{a}_A:=\int_0^\infty d{\omega} f(\omega)\,\hat{a}^\text{I}_\omega,\quad
	\int_0^\infty d{\omega}\, |f(\omega)|^2=1.
\end{align}
Using \eqref{eq:bogrindop}, this detector mode is rewritten in the form of 
\begin{align}
	\hat{a}_A=\alpha \,\hat{a}_\parallel+\beta\,\hat{a}_\perp^\dag,
\end{align}
where we defined 
\begin{align}
    \label{eq:defalpbet}
	\alpha&:=\sqrt{\int_0^\infty d\omega\, |f(\omega)|^2\,\alpha_\omega^2},
	\quad \beta:=\sqrt{\int_0^\infty d\omega \,|f(\omega)|^2\,\beta_\omega^2},\quad \alpha^2-\beta^2=1,\\
	\hat{a}_\parallel&:=\frac{1}{\alpha}\int_0^\infty d\omega {f(\omega)}\alpha_\omega\,\hat{a}^\Phi_\omega,
	\quad\hat{a}_\perp:=\frac{1}{\beta}\int_0^\infty d\omega f^*(\omega)\beta_\omega\,\hat{a}^\Psi_\omega.
\end{align}
From the partner formula \eqref{eq:partnerformula}, the partner mode P for mode A is
\begin{align}
	\hat{a}_P=\alpha \,\hat{a}_\perp+\beta\,\hat{a}_\parallel^\dag.
\end{align}
This partner mode can be reformulated using Rindler modes and Milne modes:
\begin{align}
	\hat{a}_P=\int_0^\infty d{\omega}\, \frac{\alpha_\omega\beta_\omega}{\alpha\beta}f^*(\omega)\,\hat{a}^\text{II}_\omega+\int_0^\infty d{\omega}\, \frac{\beta^2-\beta_\omega^2}{\alpha\beta}f^*(\omega)\left(\hat{a}^\text{I}_\omega\right)^\dag.
	\label{eq:partnerrindo}
\end{align}
This is a key equation in this section. 
The partner mode of the superposed Rindler modes is a superposition of the Milne modes and the Rindler modes with squeezing.
For the single mode case,  the weighting function is $|f(\omega)|^2=\delta(\omega-\omega_0)$, and $\beta=\beta_{\omega_0}$ holds. Thus, the contribution of the Rindler modes in  Eq. \eqref{eq:partnerrindo} vanishes, and the spatial profile of the partner mode is a mirror-reflection  of mode A's profile  with respect to the Rindler horizon $V=0$.
On the other hand, in general, the superposition of modes results in the second term in Eq. \eqref{eq:partnerrindo} not being equal to zero. Typically, the partner of the Rindler modes is not restricted to region II and extends beyond the Rindler horizon. This leakage can be explicitly observed by analyzing the profile function of these modes.

The general form of detector modes in Minkowski spacetime is given as follows:
\begin{align}
    \label{eq:arbitrindmiln}
	\hat{a}:=\int_0^\infty d\omega \left\{A(\omega)\,\hat{a}^\text{I}_\omega+B(\omega)\left(\hat{a}^\text{I}_\omega\right)^\dag+C(\omega)\,\hat{a}^\text{II}_\omega+D(\omega)\left(\hat{a}^\text{II}_\omega\right)^\dag\right\},
\end{align}
where $A(\omega)$, $B({\omega})$, $C(\omega)$ and $D(\omega)$ are weighting functions.
On the other hand, for a massless (chiral) scalar field, it is also possible to define the detector mode by smearing the field operator $\hat{\Pi}(V)=\partial_V\hat\phi(V)$ \cite{PhysRevD.107.085002,tomitsuka2020partner,osawa2024final}:
\begin{align}
	\hat{Q}&:=\int d{V} q(V)\,\hat{\Pi}(V),\quad
	\hat{P}:=\int d{V} p(V)\,\hat{\Pi}(V),
    \label{eq:APi}
\end{align}
where the profile functions $q(V)$ and $p(V)$ of the mode are real functions chosen to satisfy the canonical commutation relation
\begin{align}
	[\hat{Q},\hat{P}]=-\frac{i}{2}\int d{V} q(V)\,p'(V)\equiv i.
\end{align}
By expanding the field operator in terms of annihilation operators and comparing it with Eq.~\eqref{eq:arbitrindmiln}, we obtain the profile functions of the detector modes expressed in terms of these weighting functions (See Appendix~\ref{sec:app-derprof} for detailed derivations):
\begin{align}
    \label{eq:windowrepq}
	q(V)&=-2\sqrt{2}\,\Im\left[\int_0^\infty d\omega \left\{\left(A(\omega)+B^*(\omega)\right)\left(\phi_\omega^\text{I}(V)\right)^*+\left(C(\omega)+D^*(\omega)\right)\left(\phi_{-\omega}^\text{II}(V)\right)^*\right\}\right],\\
    \label{eq:windowrepp}
	p(V)&=2\sqrt{2}\,\Re\left[\int_0^\infty d\omega \left\{\left(A(\omega)-B^*(\omega)\right)\left(\phi_\omega^\text{I}(V)\right)^*+\left(C(\omega)-D^*(\omega)\right)\left(\phi_{-\omega}^\text{II}(V)\right)^*\right\}\right].
\end{align}
\begin{figure}[H]
	\centering
        \includegraphics[width=0.9\linewidth]{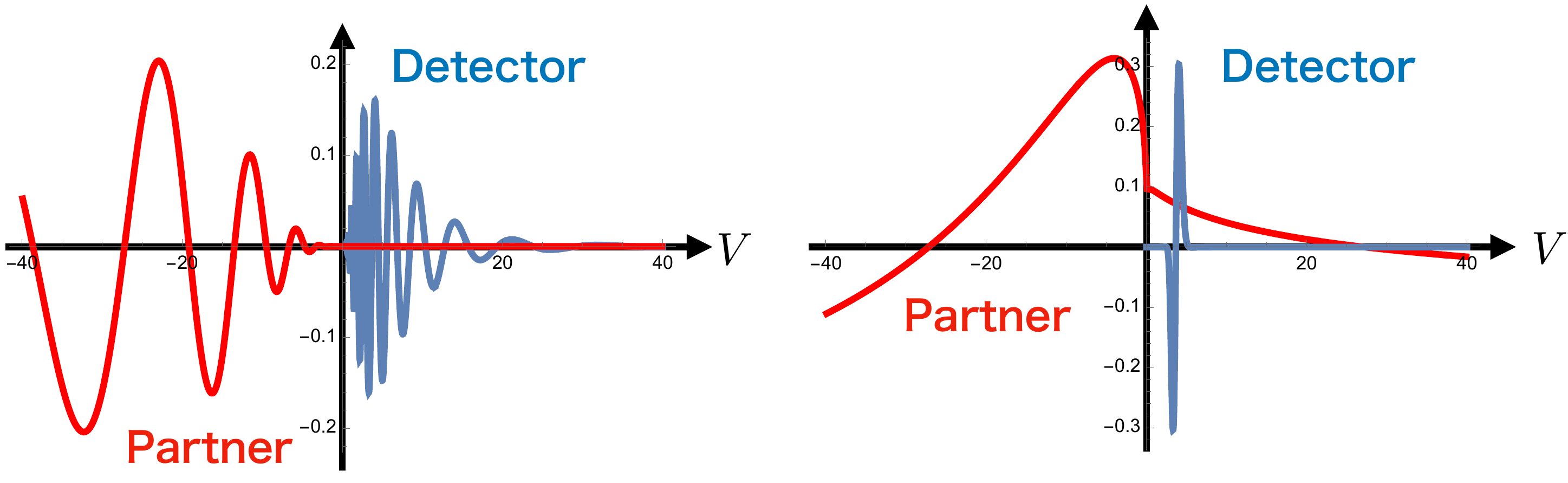}
        \caption{The profile function $q_A(V)$ of the detector mode A {placed in region I ($V>0$). Mode A }is defined by $\hat{a}_A:=\int_0^\infty d\omega \,\omega^{3/2}e^{-\alpha (\omega-\omega_0)^2}\hat{a}^\text{I}_\omega$ (blue lines), and its partner mode $q_P(V)$ (red lines). The left panel shows the profile functions with $\alpha=50$, $\omega_0=1.5$. The right panel shows the profile functions with $\alpha=1$, $\omega_0=0.05$. In these plots, we choose $a=0.1$ as the acceleration of the Rindler observer.}
        \label{fig:overlapped}
\end{figure}
Profiles of detector mode A and its partner mode are shown in Fig.~\ref{fig:overlapped}. In the left panel, the weighting function $f(\omega)$ of the Rindler mode is sharply localized in the frequency domain. As a result, the detector mode A behaves as a mode with a single frequency (single mode), and 
its partner mode is located on the opposite side of the Rindler horizon $V=0$. However, due to the factor $\alpha_\omega\beta_\omega$  in Eq.~\eqref{eq:partnerrindo} that contributes as a shift  of the weighting function toward  smaller frequencies, the partner wave packet is not the mirror-flipped image of the detector mode A but becomes the redshifted one. To reduce this redshift, we need to consider a more sharply localized weighting function in the frequency domain ($\alpha\gg1/a$); however, the wave packet is not localized in the configuration space in such a case. It is worth noting that the detector mode coincides with the Rindler mode in this limit.
The right panel of Fig.~\ref{fig:overlapped} shows the profile functions for the case in which the single-mode approximation is not applicable. We observe that the profile  of the partner mode is not localized in region II, but leaks out from the Rindler horizon in this case. 

Now let us prepare another detector mode B, which consists of the superposition of Rindler modes orthogonal to mode A:
\begin{align}
	\hat{a}_B:=\int_0^\infty d{\omega} \,g(\omega)\,\hat{a}^\text{I}_\omega.
    \label{eq:defmodeBF}
\end{align}
The weighting function $g(\omega)$ satisfies the following normalization condition and the orthogonality condition for the weighting function $f(\omega)$:
\begin{align}
	\int_0^\infty d\omega |g(\omega)|^2=1,\quad
	\int_0^\infty d\omega f(\omega)g^*(\omega)=0.
\end{align}
If we choose the weighting function $g(\omega)$ properly (for example, we take $g(\omega)\propto(\beta^2-\beta_\omega^2) f(\omega)$ \footnote{Recall the definition of $\beta$ in Eq.~(\ref{eq:defalpbet}) to verify the independecy of $\hat{a}_A$ and $\hat{a}_B$.} which comes from the second term of Eq.~(\ref{eq:partnerrindo})), the profile function of $\hat{a}_B$ and the profile function of the partner mode $\hat{a}_P$ of $\hat{a}_A$ overlap, and we expect the success of entanglement harvesting from the quantum field using the bipartite detector mode AB. 
However, it can be shown that harvesting with the mode AB fails by investigating the discriminant $D$ directly:
From Eq.~(\ref{eq:partnerrindo}), the partner mode $\hat{a}_P$ of  the superposed Rindler mode does not contain the annihilation part $\hat{a}^\text{I}_\omega$ of the Rindler mode.
As a result, the following relation holds for $\hat{a}_B$  defined as Eq.~(\ref{eq:defmodeBF}),
\begin{align}
		[\hat{a}_P^\dag,\hat{a}_B]=0,
\end{align}
and the discriminant $D$ becomes greater than zero. 
Therefore, by Simon's criterion, the detector modes A and B do not entangle, and the entanglement harvesting using the two detector modes A and B does not succeed.

We summarize the result of this section as the following theorem:
\begin{thm}[No-go theorem of harvesting]
 It is impossible to extract entanglement from a chiral quantum field using detector modes A and B when their annihilation operators $\hat{a}_A$ and $\hat{a}_B$ consist solely of Rindler annihilation operators:
 \begin{align}
    \label{eq:nogoeq}
    \hat{a}_A=\int_0^\infty d{\omega} f(\omega)\,\hat{a}^{\mathrm{I}}_\omega, \quad
    \hat{a}_B=\int_0^\infty d{\omega}\, g(\omega)\,\hat{a}^{\mathrm{I}}_\omega.
 \end{align}
\end{thm}
This no-go theorem also holds for the Milne modes, which can be verified by replacing the Rindler modes with the Milne modes.
Considering detector modes constructed from Minkowski modes provides a useful perspective for understanding our no-go theorem. Specifically, when the Minkowski vacuum is reduced by tracing out all degrees of freedom except for those of a detector mode composed purely of annihilation operators, the resulting state remains pure. Consequently, a pair of such detector modes cannot extract quantum entanglement. This illustrates that our no-go theorem can be regarded as a generalization of this Minkowski-space result.

In an accelerating frame, two independent modes—Rindler or Milne modes—emerge as a result of the horizon, and these modes are strongly entangled with one another. Since the reduced state of these Rindler and Milne modes remains nearly pure, a pair of detector modes localized on the same side of the horizon is unable to extract entanglement, even in the presence of partner-mode leakage across the horizon. This perspective may be further deepened by invoking Camalet-type monogamy relations \cite{PhysRevLett.119.110503,PhysRevResearch.2.043068,nambu2023entanglement}, which limit the entanglement between two detector modes (internal correlation) based on the external  entanglement in which these bipartite modes participate.

\subsection{The case no-go theorem is not applicable}
In this subsection, we explore a case where our no-go theorem is not applicable and analyze the resulting physical implications. The no-go theorem loses its applicability when the detector mode $\hat{a}_{B'}$ is utilized, incorporating the creation component
	\begin{align}
        \label{eq:rindwsq}
		\hat{a}_{B'}=\int_0^\infty d\omega \left\{g(\omega)\,\hat{a}^{\text{I}}_\omega+h(\omega)\left(\hat{a}^{\text{I}}_\omega\right)^\dag\right\},
	\end{align}
as the commutator $\left[\hat{a}_P^\dag,\hat{a}_{B'}\right]$ does not vanish. Therefore, employing the detector mode $\hat{a}_A$, which includes only the annihilation operator $\hat a_\omega^\text{I}$, alongside $\hat{a}_{B'}$ as specified by \eqref{eq:rindwsq}, allows for entanglement extraction without restrictions.

To observe the difference between $\hat{a}_B$ and $\hat{a}_{B'}$, we consider two UDW detectors whose time evolution operators are given by
\begin{align}
    \hat{U}_B&=1-i\lambda\left(\hat{\sigma}_+\,\hat{a}_B+\hat{\sigma}_-\,\hat{a}_B^\dag\right),\\
    \hat{U}_{B'}&=1-i\lambda\left(\hat{\sigma}_+\,\hat{a}_{B'}+\hat{\sigma}_-\,\hat{a}_{B'}^\dag\right).
\end{align}
Now, we apply these time evolution operators to the quantum state $\left|0\right\rangle_D\otimes\left|0\right\rangle_{\text{I}}$ where $|0\rangle_D$ is the lowest energy state of the detector (i.e., $\hat{\sigma}_-|0\rangle_D=0$) and $|0\rangle_\text{I}$ is the Rindler vacuum state:
\begin{align}
    &\hat{U}_B\left|0\right\rangle_D\otimes\left|0\right\rangle_{\text{I}}=|0\rangle_D\otimes\left|0\right\rangle_{\text{I}},\\
    &\hat{U}_{B'}|0\rangle_D\otimes|0\rangle_{\text{I}}=|0\rangle_D\otimes|0\rangle_{\text{I}}-i\lambda\int_0^\infty d{\omega}\,h(\omega)|1\rangle_D\otimes|1_\omega\rangle_{\text{I}}.
\end{align}
On the other hand, when these time evolution operators act on the quantum state with 1-Rindler particle $|0\rangle_D\otimes|1_\Omega\rangle_{\text{I}}$, the time-evolved states are
\begin{align}
    &\hat{U}_B\left|0\right\rangle_D\otimes\left|1_\Omega\right\rangle_{\text{I}}=|0\rangle_D\otimes|1_\Omega\rangle_{\text{I}}-i\lambda \,g(\Omega)|1\rangle_D\otimes\left|0\right\rangle_{\text{I}},\\
  &\hat{U}_{B'}|0\rangle_D\otimes|1_\Omega\rangle_{\text{I}}=|0\rangle_D\otimes|1_\Omega\rangle_{\text{I}}-i\lambda \,g(\Omega)|1\rangle_D\otimes\left|0\right\rangle_{\text{I}}-i\lambda \int_0^\infty d\omega \,h(\omega)|1\rangle_D\otimes|1_\omega1_\Omega\rangle_{\text{I}}.
\end{align} 
The time evolution by $\hat{U}_B$ does not excite the detector in the absence of Rindler particles, whereas it does excite the detector with some probability if Rindler particles are present. In contrast, the time evolution operator $\hat{U}_{B'}$ can excite the detector with a non-zero probability even if no Rindler particles are present. 

Hence, we can say that the UDW detector with time evolution $\hat{U}_B$  observes the ``real" Rindler particles through its excitation. Then, what does the detector corresponding to $\hat{a}_{B'}$ observe? To investigate this, we introduce the total particle number $\hat{N}$, defined as
\begin{align}
    \hat{N}=\hat{N}_D+\int_0^\infty d{\omega} \hat{N}_\omega^\text{I},
\end{align}
where $\hat{N}_D:=\hat{\sigma}_+\hat{\sigma}_-$ is the number operator for the detector, and $\left\{\hat{N}_\omega^{\text{I}}\right\}$ is the number operator for the Rindler modes. We observe that the expectation value of $\hat{N}$ does not change under the time evolution $\hat{U}_B$, while it increases under the time evolution $\hat{U}_{B'}$. Thus, the interaction including $\hat{a}_{B'}$ performs a measurement that is accompanied by field excitation, and its measurement outcomes contain information about ``virtual" Rindler particles.
\subsection{Application to Hawking radiation}
According to the considerations on Unruh's shell collapsing model  \cite{unruh1976notes,birrell1984quantum} and the moving mirror model \cite{carlitz1987reflections,hotta2015partner,wald2019particle,osawa2024final}, the Milne modes $\hat{a}_\omega^{\text{II}}$ are transformed into the Minkowski modes $\hat{a}_\omega^{\text{BH}}$ owing to the structure of spacetimes, which gives rise to Hawking radiation as the emission of real particles. Therefore, the measurement of the Minkowski modes in black hole spacetimes is equivalent to the measurement of the Milne modes in Minkowski spacetime. Considering the relation between the Rindler modes and the Milne modes, the operators $\hat{a}_A$, $\hat{a}_B$, $\hat{a}_{B'}$ in Eqs.~(\ref{eq:nogoeq}) and (\ref{eq:rindwsq}) have the following correspondences:
\begin{align}
    \hat{a}_A\to\int_0^\infty d{\omega} f(\omega)\,\hat{a}^{\mathrm{\text{BH}}}_\omega, \quad
    \hat{a}_B\to\int_0^\infty d{\omega} \,g(\omega)\,\hat{a}^{\mathrm{\text{BH}}}_\omega,
   \quad\hat{a}_{B'}\to\int_0^\infty d\omega \left\{g(\omega)\hat{a}^{\text{BH}}_\omega+h(\omega)\left(\hat{a}^{\text{BH}}_\omega\right)^\dag\right\}.
\end{align}
Under these correspondences, $\hat{a}_{A,B}$ represents the measurement of the real particles (Hawking radiation) by the observer at the future null infinity, whereas $\hat{a}_{B'}$ corresponds to the measurements that include contributions from ``virtual particles", or in other words, vacuum fluctuations. 
For the vacuum fluctuation scenario \cite{wilczek1993quantum,carlitz1987reflections,hotta2015partner} of black hole evaporation, our no-go theorem is applicable, and the following corollary holds:
\begin{cor}[No-go theorem for vacuum fluctuation scenario]
 Extracting quantum correlations from Hawking radiation that reaches future null infinity is not feasible \textbf{when considering solely real Hawking particles originating from Milne particles}. 
\end{cor}
We can also apply our no-go theorem to Page's scenario \cite{page1993information} of black hole evaporation, in which the partner of the early stage Hawking radiation is identified with the late-time Hawking radiation. It is important to note that the correspondence between the Milne modes $\hat{a}^{\text{II}}_\omega$ and the Minkowski modes $\hat{a}^{\text{BH}}_\omega$ holds only during the early stage of black hole evaporation in this case, as the late-time radiation corresponds to the Rindler modes $\hat{a}^{\text{I}}_\omega$. Thus, above corollary also applies to Page's scenario, restricted to the early stage radiation.
%
It is important to emphasize again that this no-go theorem does not forbid entanglement extraction by using vacuum fluctuations of the quantum field in black hole spacetimes.

\subsection{Demonstrations}\label{sec:demo}
In this subsection, we numerically evaluate the entanglement between two detector modes consisting of the Milne modes in region $V<0$, as a demonstration of our no-go theorem. For comparison, we consider two different types of detector modes: a detector mode with $\omega$-top-hat profile functions (type 1) and a detector mode with sin-cos profile functions (type 2). The key difference between them lies in the structure  of the local operators represented by annihilation operators: the former does not include the creation operators of the Milne modes, whereas the latter does. As a result, our no-go theorem does not apply to the latter detector. 

The $\omega$-top-hat function detector modes $\left\{\hat{q}_A^{(1)},\hat{p}_A^{(1)},\hat{q}_B^{(1)},\hat{p}_B^{(1)}\right\}$ are defined using annihilation operators of the Milne modes as
\begin{align}
    \hat{a}_{A,B}^{(1)}=\int_{0}^{\infty} d{\omega} F_{A,B}^{(1)}(\omega)\,\hat{a}_\omega^{\text{II}},\quad  \int_0^\infty  d{\omega} \,|F_{A,B}^{(1)}(\omega)|^2=1,
\end{align}
where $F_{A,B}^{(1)}$ is the weighting function given as
\begin{align}
    F_{A,B}^{(1)}(\omega)=
    \begin{cases}
     \dfrac{e^{i \omega f(V_{A,B})}}{\sqrt{\omega_+ -\omega_-}}& \quad \text{for}\quad\omega_-\le\omega\le \omega_+\\
    0&\quad \text{else}
    \end{cases}
    \label{eq:FAB1}
\end{align}
with $0\le \omega_-<\omega_+$. Here, the function $f(V)$ is defined as $f(V)=-(1/a)\log(-a V)$.
The two detector modes become independent and satisfy the canonical commutation relations $[\hat{q}_j^{(1)},\hat{p}_k^{(1)}] =i \delta_{jk},~j,k=A,B$ 
if $V_A$ and $V_B$ satisfy
\begin{equation}
\frac{V_B}{V_A }=\exp\left[\frac{2\pi n a}{\omega_+-\omega_-}\right],\ \ n = \pm 1,\pm  2, \dots.
\label{eq:discretization}
\end{equation}   
The profile function corresponding to the weighting function  \eqref{eq:FAB1} is obtained using Eqs.~(\ref{eq:windowrepq}) and (\ref{eq:windowrepp}):
\begin{align*}
    q_{A,B}^{(1)}(V)&=-\Im\left[\left(\frac{-2i\left(f(V)-f(V_{A,B})\right)}{\pi(\omega_+-\omega_{-})}\right)^{1/2}\left(\Gamma\left(\frac{1}{2},-i(f(V)+f(V_{A,B}))\omega_-\right)-\Gamma\left(\frac{1}{2},-i(f(V)+f(V_{A,B}))\omega_+\right)\right)\right],\\
    p_{A,B}^{(1)}(V)&=\Re\left[\left(\frac{-2i\left(f(V)-f(V_{A,B})\right)}{\pi(\omega_+-\omega_{-})}\right)^{1/2}\left(\Gamma\left(\frac{1}{2},-i(f(V)+f(V_{A,B}))\omega_-\right)-\Gamma\left(\frac{1}{2},-i(f(V)+f(V_{A,B}))\omega_+\right)\right)\right],
\end{align*}
where $\Gamma(a,x)$ is the incomplete Gamma function defined by $
    \Gamma(a,x)=\int_{x}^\infty \dd{r}t^{a-1}e^{-t}$.
Please note that the spatial profiles of modes A and B are not compact and always have overlap, even if they satisfy the independent condition \eqref{eq:discretization}.

The sin-cos detector modes $\left\{\hat{q}_A^{(2)},\hat{p}_A^{(2)}, \hat{q}_B^{(2)},\hat{p}_B^{(2)}\right\}$ are defined by
\begin{align}
    \hat{q}_{A,B}^{(2)}=\int d{V}q_{A,B}^{(2)}(V)\,\hat{\Pi}(V), \quad
    \hat{p}_{A,B}^{(2)}=\int d{V}p_{A,B}^{(2)}(V)\,\hat{\Pi}(V),
\end{align}
where $q_{A,B}^{(2)}(V)$ and $p_{A,B}^{(2)}(V)$ are the profile functions, given by
\begin{align}
    q_{A,B}^{(2)}(V)=\frac{2}{\sqrt{{\pi}}}\cos(2\pi (f(V)-f(V_{A,B}))/L), \quad
     p_{A,B}^{(2)}(V)=\frac{2}{\sqrt{\pi}}\sin(2\pi (f(V)-f(V_{A,B}))/L).
\end{align}
within the region $-L/2\le f(V)-f(V_{A,B})\le L/2$, and zero otherwise.
The detector modes correspond to a superposition of the Milne modes, as the dependence on $V$ appears solely through the function $f(V)$.
When $|f(V_A)-f(V_B)|>L/2$ is satisfied, the spatial profiles of modes A and B do not overlap, and the two detector modes are independent.
In this case, the corresponding canonical operators satisfy the canonical commutation relations $[\hat{q}_j^{(2)},\hat{p}_k^{(2)}]=i\delta_{jk},~j,k=A,B$ 
which can be verified by changing the integration variable to $U=f(V)$.
We can also represent these local modes with the creation and annihilation operators of the Milne modes:
\begin{align}
    \hat{a}_{A,B}^{(2)}=\frac{\sqrt{2}i}{\pi}\int_0^\infty d\omega \sqrt{\omega}\left(\cos\frac{L\omega}{2}\right)\left(\frac{e^{-i\omega f^{-1}(V_{A,B})}}{\omega-\pi/L}\,\hat{a}_\omega^{\text{II}}
    +\frac{e^{i\omega f^{-1}(V_{A,B})}}{\omega+\pi/L}\left({\hat{a}_\omega^{\text{II}}}\right)^\dag\right).
\end{align}
It is worth noting that the entanglement entropy between the detector mode A and its complementary mode does not converge due to the UV divergence of the quantum field. In fact, the partner mode constructed using the partner formula is also ill-defined as a result of this divergence. To regularize this divergence, we should modify the definition of the detector mode as follows:
\begin{align}
    \label{eq:sin-cos-reg}
     \hat{a}_{A,B}^{(2)}=\frac{\sqrt{2}i}{\pi}\int_0^\infty d\omega \sqrt{\omega}\left(\cos\frac{L\omega}{2}\right)\left(\frac{e^{-i\omega f^{-1}(V_{A,B})}e^{-\omega\varepsilon}}{\omega-\pi/L}\,\hat{a}_\omega^{\text{II}}
    +\frac{e^{i\omega f^{-1}(V_{A,B})}e^{-\omega\varepsilon}}{\omega+\pi/L}\left({\hat{a}_\omega^{\text{II}}}\right)^\dag\right),
\end{align}
where $\varepsilon>0$ is a UV regulator for the quantum field. When $L\gg\varepsilon$, the detector modes A and B are approximately independent for $|f(V_A)-f(V_B)|>L/2$.  Of course, the original sin-cos detector modes can be recovered by taking the limit $\varepsilon/L\to 0$.

\begin{figure}[H]
    \centering
    \includegraphics[width=0.9\columnwidth]{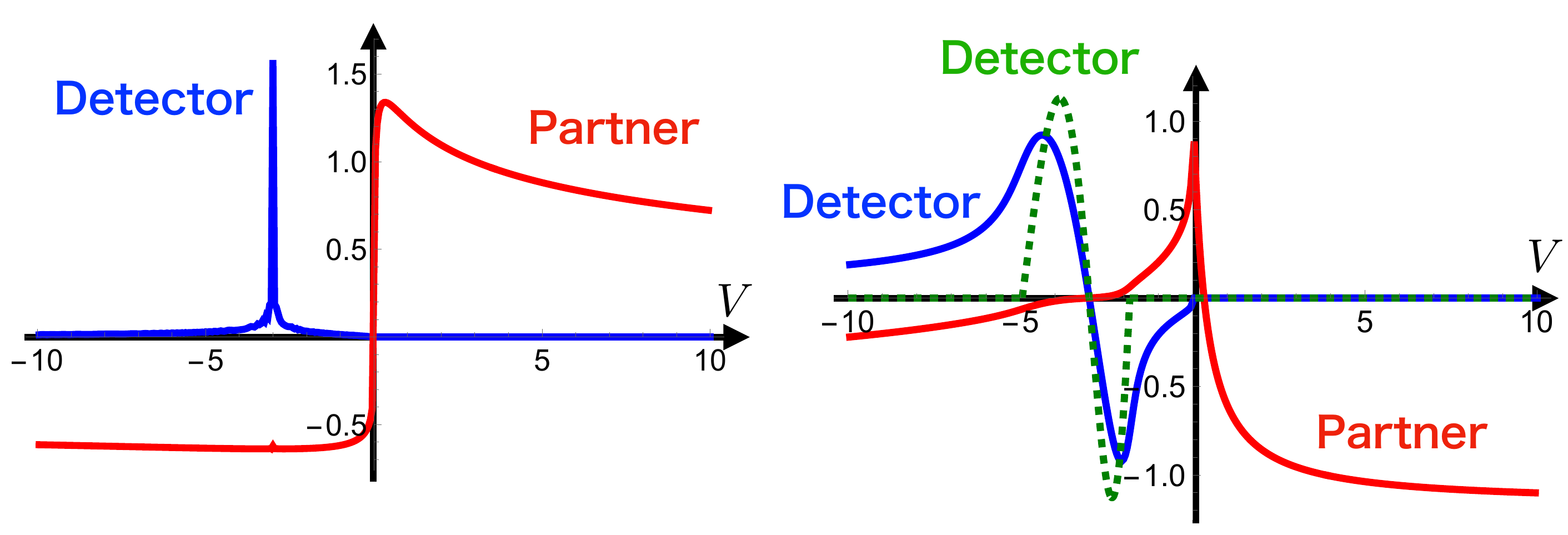}
    \caption{The left panel shows the profile functions $p_A^{(1)}(V)$ of detector modes (blue) and its partner mode $p_P^{(1)}(V)$ (red) with $a=1,\omega_-=0.1,\omega_+=1000,V_A=-3$. The right panel shows the profile functions $p_A^{(2)}(V)$ of detector modes  (blue) and its partner mode  $p_P^{(2)}(V)$ (red) with  $L=1$, $\varepsilon=0.138$, $V_A=-3$. The green dashed line is the profile function of the detector mode with $\varepsilon=0$. Note that the partner mode is ill-defined with $\varepsilon=0$.}
    \label{fig.p}
\end{figure}

The profile functions of the detector modes $\{q_A^{(1)},p_A^{(1)}\}$ (left panel) and $\{q_A^{(2)},p_A^{(2)}\}$ (right panel) are shown in Fig.~\ref{fig.p} along with their partner modes
\footnote{We choose $\varepsilon=0.138$ for the profile plot. For the sin-cos type detector mode, we plot an approximate profile in which the creation part in Eq.~\eqref{eq:sin-cos-reg} is neglected to make it easier to draw the partner profile. The ratio of the contribution from the annihilation part to that from the creation part 
\begin{align} 
    \left(\int_0^\infty d{\omega}\, \omega\cos^2 (L\omega/2)\frac{e^{-2\omega\epsilon}}{(\omega-\pi/L)^2}\right)
    \left/ \left(\int_0^\infty d{\omega}\, \omega\cos^2(L\omega/2)\frac{e^{-2\omega\epsilon}}{(\omega+\pi/L)^2}\right)\right.,
\end{align}
 reaches its maximum value $\approx 14.5$ when $\epsilon/L\approx0.138$. There, we neglected the creation part in \eqref{eq:sin-cos-reg} to make the plot.}. 
The blue lines represent the detector modes prepared in region II ($V<0$), while the red lines indicate their partner modes. The green dashed line in the right panel shows the profile function for $\varepsilon=0$. Recall that the partner formula becomes ill-defined for $\varepsilon=0$ due to UV divergence. Hence, while it is feasible to depict the detector mode profile with $\varepsilon=0$, the corresponding partner profile cannot be represented in a plot. The profile functions of the partner modes take nonzero values even in the region $V<0$ in both cases, indicating that the partner modes are not confined to one side of the horizon $V>0$.

\begin{figure}[H]
\centering
    \includegraphics[width=0.45\columnwidth]{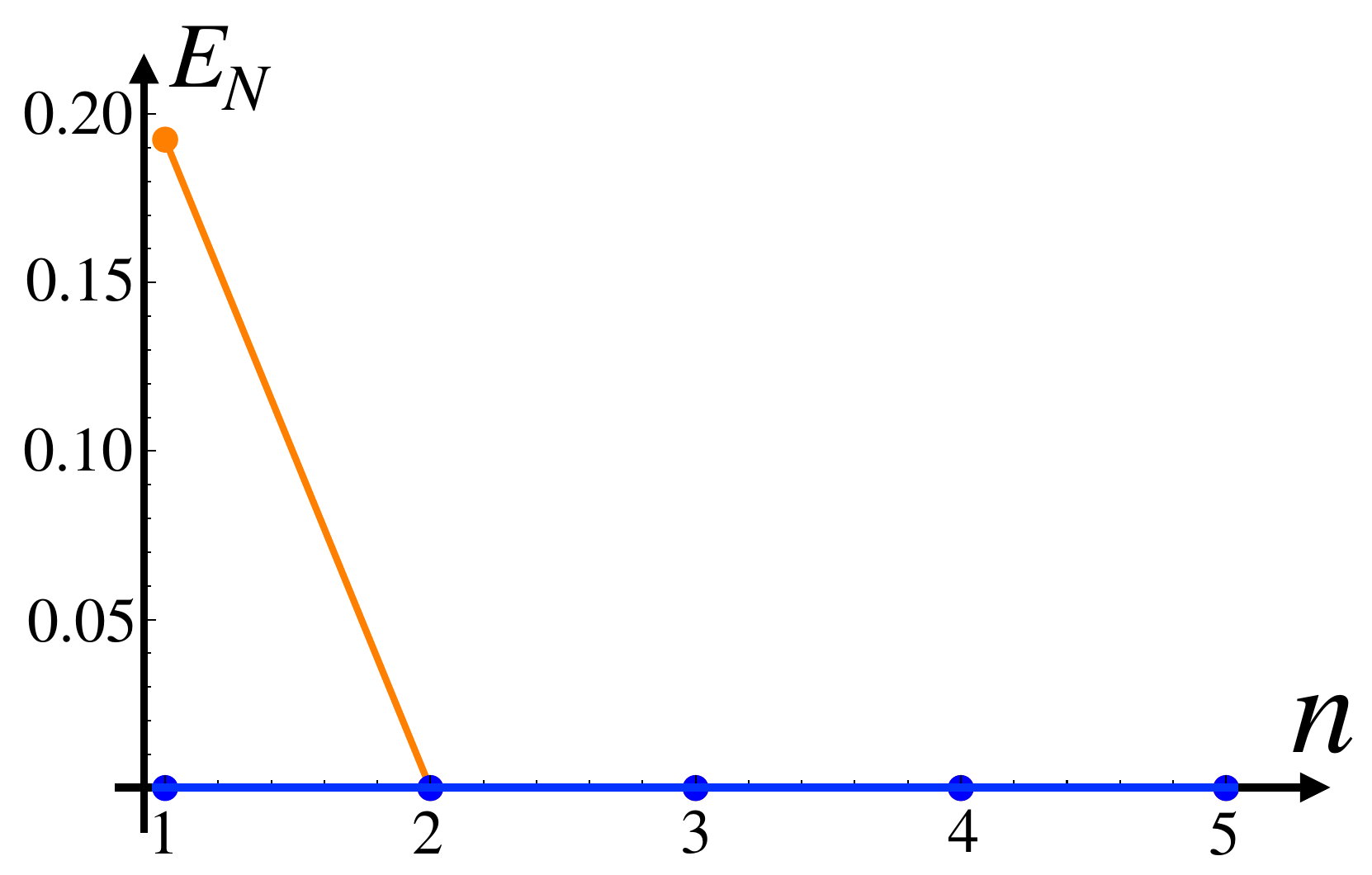}
\caption{Logarithmic negativity with  type 1  detectors ($\omega$-top-hat weighting function)  (blue) and with  type 2 detectors (the sin-cos profile) (orange). Parameters are $a=1,\omega_-=0.1,\omega_+=50,V_A=-1,L=2\pi /(\omega_+-\omega_-)$, and $\varepsilon=0.02$.}
\label{fig.a}
\end{figure}

We use logarithmic negativity $E_N$ \cite{peres1996separability,horodecki1997separability,simon2000peres} to quantify the entanglement between mode A and mode B:
\begin{equation}
    E_N :=-\text{min}[\log_2 (2\tilde{\nu}_-),0],
\end{equation}
where $\tilde{\nu}_-$ is the minimum symplectic eigenvalue of the partially transposed covariance matrix of the bipartite state AB.
In Fig.~\ref{fig.a}, we plot the negativity between local modes as a function of $n$. The integer $n$ determines the separation between local modes 
according to Eq.~(\ref{eq:discretization}).
The blue line represents entanglement between type 1 detectors  (with the $\omega$-top-hat weighting function  modes $\{\hat{q}_{A,B}^{(1)},\hat{p}_{A,B}^{(1)}\}$), while the orange line represents entanglement between type 2 detectors (the sin-cos type  modes $\{\hat{q}_{A,B}^{(2)}, \hat{p}_{A,B}^{(2)}\}$).
For type 1 detector modes, two detector modes remain separable regardless of their separation, consistent with the general discussion based on the discriminant $D$. In contrast, type 2  detector modes allow entanglement extraction for $n=1$. This occurs as the type 2 detector modes incorporate the creation operator of the Milne mode, rendering our no-go theorem irrelevant.
The conditions for successful entanglement extraction are revisited from the perspective of the shape of the partner mode. As shown in Sec.~\ref{sec:profilerep}, the discriminant $D$ is expressed in terms of the commutators between the canonical operator of the partner  mode and the detector mode B. For the chiral field, these commutators can be rewritten in terms of the products of the derivative of the partner profile function and the profile function of the detector mode B, and the discriminant is
\begin{align}
    D&=\frac{1}{2}\int dx\, dy \,q'_P(x)p'_P(y)\left(-q_B(x)p_B(y)
    +p_B(x)q_B(y)\right)
    .
\end{align}
If the detector mode B is placed at the location where the derivative of the partner profile $q_P(x)$ or $p_P(x)$ vanishes, then entanglement extraction fails because $D=0$. This expectation is consistent with the partner profiles shown in Fig.~\ref{fig.p} and the behavior of the negativity in Fig.~\ref{fig.a}. Actually, regarding the location of the detector (the center of the detector profile), the profile of the partner is flat, and its derivative becomes zero (left panel of Fig. \ref{fig.p}), whereas the derivative of the partner profile is nonzero in the right panel of Fig. \ref{fig.p}, in which case entanglement harvesting is possible.
\section{Conclusion} \label{sec:conclusion}
We revisited Simon's criterion of entanglement \cite{simon2000peres} for the bipartite Gaussian states from the perspective of the partner mode by Hotta \textit{et al}.~\cite{hotta2015partner}. 
Employing this reformulated Simon's criterion, we observed that the overlap between the spatial profiles of the detector mode B and the spatial profiles (the derivative of the spatial profiles for the chiral field) of the partner  of the detector mode A is necessary for establishing entanglement between A and B.
Thus, our clear understanding of the partner mode's profile—described as ``the profile function of the partner mode indicates the location of the complementary information of the detector mode's entanglement"—is indeed accurate.

It is important to note that the overlap of profile functions does not serve as an equivalent condition to Simon's criterion. In certain instances, Simon's criterion can dismiss the potential for entanglement even when the condition of overlap is fulfilled.
An instance of such scenarios is the extraction of entanglement utilizing detectors composed of superpositions of the Rindler modes in region I. The partner mode of a single Rindler mode in region I is the Milne mode in region II. Hence, the location of the partner mode is separated from that of the Rindler mode. On the other hand, if we consider the superposition of the Rindler modes in region I, the partner mode is no longer confined to region II and leaks out from the horizon. 
In this case, another detector mode $\hat{a}_B$ defined as the superposition of the Rindler modes in region I can have overlap with the partner mode $\hat{a}_{P}$ of the first detector mode $\hat{a}_A$, and entanglement harvesting is expected to be possible by using the detector mode consisting of the Rindler modes in region I. 
However, our no-go theorem forbids
this entanglement harvesting when $\hat{a}_A$ and $\hat{a}_B$ consist solely of the annihilation operators of the Rindler modes. In contrast, the theorem does not prohibit entanglement harvesting using a detector mode $\hat{a}_{B'}$ containing a creation part that corresponds to the squeezing of the mode.
Such squeezing can arise in the case that  the detector B$'$ accelerates with an acceleration different from that of the Rindler detector A.
The entangling/disentangling phenomena between the accelerating detector and the inertial detector \cite{bruschi2010unruh,lin2008disentanglement} may be one example.

The relationship between the Unruh effect and Hawking radiation can be explored through the moving mirror model \cite{carlitz1987reflections}, which mimics Unruh's shell collapsing model of black hole formation \cite{unruh1976notes}. 
In this framework, it has been shown that Hawking radiation originates from the Milne mode at past null infinity, reflecting off the mirror and ultimately transforming into real particle modes \cite{hotta2015partner,wald2019particle,osawa2024final}.  
Hence, by interchanging the roles of the Rindler modes and the Milne modes, the Unruh effect is directly related to Hawking radiation. 
The detector modes that detect Hawking radiation (real particles at future null infinity) correspond to detector modes consisting of positive frequency Milne modes in past null infinity.
By applying the partner formula for the Rindler modes Eq.~\eqref{eq:partnerrindo} to Hawking radiation in  black hole spacetimes, we find that the partner of the Hawking radiation is not necessarily confined inside the black hole.
This scenario has already been pointed out by \cite{hotta2015partner}.

How should we interpret our no-go theorem in the context of black hole evaporation? By applying our
no-go theorem to the Milne modes, we find that harvesting is impossible when the detector modes consist solely of Milne mode annihilation operators.
Since these detector modes correspond to measurements of real particles emitted from a black hole via Hawking radiation, our result implies that no quantum correlations exist between real particles emitted as Hawking radiation if we follow the vacuum fluctuation scenario \cite{wilczek1993quantum,carlitz1987reflections,hotta2015partner}.
Even under Page's evaporation scenario \cite{page1993information}, quantum correlations among (real) Hawking radiation are absent during the early stages of black hole evaporation. 
It is important to note that vacuum fluctuations do restore quantum correlations even in the early stages of black hole evaporation, as discussed previously. These correlations can be extracted by exciting the vacuum fluctuation, for example, through the measurement of the localized field (as discussed in Sec.~\ref{sec:demo}), by using accelerating detectors, or by employing the spatial superposition of detectors \cite{foo2020unruh,barbado2020unruh}.
However, for an observer waiting and observing Hawking radiation at infinity, it is impossible to extract entanglement by measuring real Hawking particles that reach him.

\begin{acknowledgements}
Y.O. would like to take this opportunity to thank the “Nagoya University Interdisciplinary Frontier Fellowship” supported by Nagoya University and JST, the establishment of university fellowships towards the creation of science technology innovation, Grant Number JPMJFS2120. Y. N. was supported by JSPS KAKENHI (Grant No. 23K25871) and MEXT KAKENHI Grant-in-Aid for Transformative Research Areas A “Extreme Universe” (Grant No. 24H00956).
\end{acknowledgements}

\appendix
\section{Introduction of the parametrization}\label{ap-derivation of parametrization}
Let us consider two independent oscillator modes $\hat{a}_A$ and $\hat{a}_B$ embedded in $n$-mode Gaussian modes $\{\hat{b}_1,\cdots,\hat{b}_n\}$:
\begin{align}
    &\hat{a}_A:=\sum_{i=1}^n \left\{\alpha_i^A\,\hat{b}_i+\beta_i^A(\hat{b}_i)^\dag\right\},\quad
    \hat{a}_B:=\sum_{i=1}^n \left\{\alpha_i^B\,\hat{b}_i+\beta_i^B(\hat{b}_i)^\dag\right\}, \\
&\sum_i(|\alpha_i^A|^2-|\beta_i^A|^2)=1,\quad\sum_i(|\alpha_i^B|^2-|\beta_i^B|^2)=1.
\end{align}
Without loss of generality,  either $\hat{a}_A$ or $\hat{a}_B$ 
can be rewritten in terms of the squeezing of two independent oscillator modes \cite{hotta2015partner}. 
In this paper, we transform the oscillator mode $\hat{a}_A$. $\hat{a}_A$ is rewritten as 
\begin{align}
    \label{eq:apmodeAws}
    \hat{a}_A=\alpha\,\hat{a}_{\tilde{\parallel}}+\gamma\,\hat{a}_{\tilde{\parallel}}^\dag+\delta\,\hat{a}_\perp^\dag,
\end{align}
where $\hat{a}_{\tilde{\parallel}}$ is the annihilation part of $\hat{a}_A$, and $\alpha$ is its normalization:
\begin{align}
    \hat{a}_{\tilde{\parallel}}=\frac{1}{\alpha}\sum_{i=1}^n\alpha_i^A\,\hat{b}_i,\quad
    \alpha={\sqrt{\sum_{i=1}^n|\alpha_i^A|^2}}.
\end{align}
The parameters $\gamma$, $\delta$, and the annihilation operator $\hat{a}_\perp$ are defined as
\begin{align}
    \gamma&:=[\hat{a}_{\tilde{\parallel}},\hat{a}_A]=\frac{1}{\alpha}\sum_{i=1}^n{\left(\alpha_i^A\right)\left(\beta_i^A\right)},\\
    \delta&:=\sqrt{\alpha^2-|\gamma|^2-1},\\
    \hat{a}_\perp&:=\frac{1}{\delta}\left(\sum_{i=1}^n(\beta_i^A)^*\,\hat{b}_i-\gamma^*\,\hat{a}_{\tilde{\parallel}}\right)
    =\frac{1}{\delta}\sum_{i=1}^n\left(\left(\beta_i^A\right)^*-\frac{\gamma^*\alpha_i^A}{\alpha}\right)\hat{b}_i.
\end{align}
In this procedure, we extract the component of the creation part of $\hat{a}_A$ that is orthogonal to $\hat{a}_{\tilde{\parallel}}$.

This form of $\hat{a}_A$ still contains single mode squeezing. To eliminate this, we introduce a new annihilation operator $\hat{a}_\parallel$ defined by
\begin{align}
    \hat{a}_\parallel:=
    \cosh\tilde{r}\, \hat{a}_{\tilde{\parallel}}
    -e^{i\phi} \sinh \tilde{r}\,\hat{a}^\dag_{\tilde{\parallel}},
\end{align}
where the parameters $\tilde{r}$ and $\phi$ are given by
\begin{align}
    \tanh\tilde{r}=-\frac{|\gamma|}{\alpha},\quad
    \phi=\text{Arg } \gamma.
\end{align}
By solving this relation for $\hat{a}_{\tilde{\parallel}}$ and substituting it into Eq.~(\ref{eq:apmodeAws}), we obtain
\begin{align}
    \label{eq:defmodeAap}
	\hat{a}_A=\cosh r\,\hat{a}_\parallel+\sinh r\,\hat{a}^\dag_\perp.
\end{align}
where the parameter $r$ is given by
\begin{align}
    \tanh r:=\frac{\delta}{\alpha\cosh \tilde{r}+|\gamma|\sinh\tilde{r}}.
\end{align}

We now perform a basis transformation from $\left\{\hat{b}_1,\cdots,\hat{b}_n\right\}$ to $\left\{\hat{\bar{b}}_1,\cdots,\hat{\bar{b}}_n\right\}$, constructed so that $\hat{\tilde{b}}_1=\hat{a}_\parallel$ and $\hat{\tilde{b}}_2=\hat{a}_\perp$.
In this new basis, the oscillator mode $\hat{a}_B$ can be expressed as
\begin{align}
    \hat{a}_B=\sum_{i=1}^n\left\{\bar{\alpha}_i^B\,\hat{\bar{b}}_i+\bar{\beta}_i^B\left(\hat{\bar{b}}_i\right)^\dag\right\}.
\end{align}
By introducing operators $\hat{a}_{\parallel'}$, $\hat{a}_{\perp'}$, and the parameter $r_1$, the oscillator mode $\hat{a}_B$ can be rewritten as
\begin{align}
    \hat{a}_B=\cosh r_1\,\hat{a}_{\parallel'}+\sinh r_1\,\hat{a}^\dag_{\perp'},
\end{align}
where $\hat{a}_{\parallel'},\hat{a}_{\perp'}$ are defined as
\begin{align}
    \hat{a}_{\parallel'}=\frac{\sum_{i=1}^n\bar{\alpha}^B_i\,\hat{\bar{b}}_i}{\sqrt{\sum_{i=1}^n|\bar{\alpha}^B_i|^2}},
    \quad\hat{a}_{\perp'}=\frac{\sum_{i=1}^n\left(\bar{\beta}^B_i\right)^*\hat{\bar{b}}_i}{\sqrt{\sum_{i=1}^n|\bar{\beta}^B_i|^2}}.
\end{align}
with 
\begin{equation}
\cosh r_1=\sqrt{\sum_{i=1}^n|\bar{\alpha}^B_i|^2},\quad\sinh r_1=\sqrt{\sum_{i=1}^n|\bar{\beta}^B_i|^2}.
\end{equation}
The oscillator modes $\hat{a}_{\parallel'}$, $\hat{a}_{\perp'}$ generally do not coincide with $\hat{a}_{\parallel}$, $\hat{a}_{\perp}$. Furthermore, $\hat{a}_{\parallel'}$ and  $\hat{a}_{\perp'}$ are not independent (i.e., $[\hat{a}_{\parallel'},\hat{a}_{\perp'}^\dag]\ne0$) in general.

In the following discussion, we assume that the coefficients $\bar{\alpha}_1^B$ and $\bar{\alpha}_2^B$ are real for simplicity. This assumption does not affect the results concerning quantum entanglement, as complex phase factors can be removed through local mode transformations that do not change entanglement between systems.
By using Gram-Schmidt decomposition, the oscillator modes $\hat{a}_{\parallel'}$ and $\hat{a}_{\perp'}$ can be decomposed as
\begin{align*}
    \hat{a}_{\parallel'}&=\cos\theta_1\,\hat{a}_\parallel+\sin\theta_1\cos\xi_1\,\hat{a}_\perp+\sin\theta_1\sin\xi_1\,\hat{a}_0,\\
    \hat{a}_{\perp'}&=\cos\theta_2\,\hat{a}_\parallel+\sin\theta_2\cos\xi_2\,\hat{a}_\perp+\sin\theta_2\sin\xi_2\cos\chi\,\hat{a}_0+\sin\theta_2\sin\xi_2\sin\chi\,\hat{a}_1.
\end{align*}
Here, $\hat{a}_0$ and $\hat{a}_1$ are oscillator modes independent of $\hat{a}_\parallel$ and $\hat{a}_\perp$, and they are determined as
\begin{align}
    \hat{a}_0&\propto\hat{a}_{\parallel'}-\left[\hat{a}_{\parallel'},\hat{a}_\parallel^\dag\right]\hat{a}_\parallel-\left[\hat{a}_{\parallel'},\hat{a}_\perp^\dag\right]\hat{a}_\perp,\\
     \hat{a}_1&\propto\hat{a}_{\perp'}-\left[\hat{a}_{\perp'},\hat{a}_\parallel^\dag\right]\hat{a}_\parallel-\left[\hat{a}_{\perp'},\hat{a}_\perp^\dag\right]\hat{a}_\perp-\left[\hat{a}_{\perp'},\hat{a}_0^\dag\right]\hat{a}_0.
\end{align}
The proportionality factors are determined by normalization conditions:
\begin{align}
    \left[\hat{a}_0,\hat{a}_0^\dag\right]=\left[\hat{a}_1,\hat{a}_1^\dag\right]=1.
\end{align}
Therefore, $\hat{a}_B$ can be written as a linear combination of four independent oscillator modes $\hat{a}_\parallel$, $\hat{a}_\perp$, $\hat{a}_0$, $\hat{a}_1$
\begin{align}
	\label{eq:defmodeBap}
	\hat{{a}}_B&=\cosh r_1\left(\cos\theta_1\,\hat{a}_\parallel+\sin\theta_1\cos\xi_1\,\hat{a}_\perp+\sin\theta_1\sin\xi_1\,\hat{a}_0\right)\nn
	&+\sinh r_1\left(\cos\theta_2\,\hat{a}_\parallel^\dag+\sin\theta_2\cos\xi_2\,\hat{a}_\perp^\dag+\sin\theta_2\sin\xi_2\cos\chi\,\hat{a}_0^\dag+\sin\theta_2\sin\xi_2\sin\chi\,\hat{a}_1^\dag\right).
\end{align}
\section{Proof for more general cases}\label{sec:app-generalize}
In this Appendix, we show that our entanglement criterion based on $D$ remains applicable even when the detector mode A includes single-mode squeezing:
\begin{align}
    \label{eq:modeAwithss}
    \hat{a}_A:=\cosh \tilde{r}\,\hat{a}_{\tilde{\parallel}}+\sinh \tilde{r}\cos\theta\,\hat{a}_{\tilde{\parallel}}^\dag+\sinh \tilde{r}\sin\theta\,\hat{a}_\perp^\dag.
\end{align}
As mentioned in Appendix~\ref{ap-derivation of parametrization}, the operator $\hat{a}_A$ can be rewritten as
\begin{align}
    \hat{a}_A=\cosh r\,\hat{a}_{\parallel}+\sinh r\,\hat{a}_{\perp}^\dag,
\end{align}
by performing the local symplectic transformation with an appropriate squeezing parameter $r'$:
\begin{align}
    \hat{a}_{\parallel}=\cosh r'\,\hat{a}_{\tilde{\parallel}}+\sinh r'\,\hat{a}_{\tilde{\parallel}}^\dag.
\end{align}
The partner formula, evaluated in the $\{\hat{a}_\parallel, \hat{a}_\perp\}$ basis, reads:
\begin{align}
    \hat{a}_P=\cosh r\,\hat{a}_\perp+\sinh r\,\hat{a}_\parallel^\dag.
\end{align}
Hence, the discriminant $D$ of entanglement is 
\begin{align}
    D&:=-\left|\left[\hat{a}_P,\hat{a}_B^\dag\right]\right|^2+\left|\left[\hat{a}_P,\hat{a}_B\right]\right|^2\nn
    &=\frac{-\cosh^2r\cos^2\theta_2+\sinh^2 r\sin^2\theta^2\cos^2\xi^2}{\cosh^2 r\sinh^2 r}\sinh^2 r_1,
\end{align}
as derived in Sec.~\ref{sec:condent}. In that section, we relate this discriminant $D$ to the determinant of the submatrix $V_{AB}$ of the covariance matrix evaluated for the quantum state $|0000\rangle_{\parallel\perp01}$. 

However, the detector mode $\hat{a}_A$ in Eq.~(\ref{eq:modeAwithss}) is naturally defined in terms of  annihilation operators $\{\hat{a}_{\tilde{\parallel}},\hat{a}_\perp\}$, and we must evaluate the covariance matrix for the quantum state $|0000\rangle_{\tilde{\parallel}\perp01}$.
The submatrix $\tilde{V}_{AB}$ of the covariance matrix evaluated for the quantum state $|0000\rangle_{\tilde{\parallel}\perp01}$, is given as
\begin{align}
    \tilde{V}_{AB}=\left(\begin{array}{cc}
	\left\langle\left\{\hat{q}_A,\hat{q}_B\right\}\right\rangle_{\tilde{\parallel}\perp01} &0\\
	0&\left\langle\left\{\hat{p}_A,\hat{p}_B\right\}\right\rangle_{\tilde{\parallel}\perp01}
	\end{array}\right)
\end{align}
where the expectation values are computed as
\begin{align}
    &\left\langle\left\{\hat{q}_A,\hat{q}_B\right\}\right\rangle_{\tilde{\parallel}\perp01}=\left(\cosh 2r'+\sinh 2r'+1\right)\left(\cosh r\cos\theta_2+\sinh r \sin\theta_2\cos\xi_2\right)\sinh r_1,\nn
    &\left\langle\left\{\hat{p}_A,\hat{p}_B\right\}\right\rangle_{\tilde{\parallel}\perp01}=-\left(\cosh 2r'-\sinh 2r'+1\right)\left(\cosh r\cos\theta_2-\sinh r \sin\theta_2\cos\xi_2\right)\sinh r_1.
\end{align}
Here, we used the independency conditions Eqs.~(\ref{eq:indep-1}) and (\ref{eq:indep-2}) to eliminate $\theta_1$ and $\xi_1$. Thus, the determinant of the submatrix $\tilde{V}_{AB}$ of the covariance matrix is computed as
\begin{align}
    \det \tilde{V}_{AB}&=-2(\cosh 2r'+1)\left(\cosh^2 r\cos^2\theta_2-\sinh^2 r\sin^2\theta_2\cos^2\xi_2\right)\sinh^2r_1\nn
    &=-\frac{\sinh^2 2r}{2}\left(\cosh 2r'+1\right)\left(\left|\left[\hat{a}_P,\hat{a}_B^\dag\right]\right|^2-\left|\left[\hat{a}_P,\hat{a}_B\right]\right|^2\right)\nn
    &\equiv \frac{\sinh^2 2r}{2}\left(\cosh 2r'+1\right)D.
\end{align}
Therefore, Simon's entanglement criterion $\det \tilde{V}_{AB}<0$ once again coincides with the condition $D<0$ for the discriminant $D$. 
\section{Derivation of the profile function}\label{sec:app-derprof}
In this Appendix, we will derive the following profile function representation
\begin{align}
	q(V)&=-2\sqrt{2}\Im\left[\int_0^\infty d\omega \left\{\left(A(\omega)+B^*(\omega)\right)\left(\phi_\omega^\text{I}(V)\right)^*+\left(C(\omega)+D^*(\omega)\right)\left(\phi_{-\omega}^\text{II}(V)\right)^*\right\}\right], \label{eq:B1}\\
	p(V)&=2\sqrt{2}\Re\left[\int_0^\infty d\omega \left\{\left(A(\omega)-B^*(\omega)\right)\left(\phi_\omega^\text{I}(V)\right)^*+\left(C(\omega)-D^*(\omega)\right)\left(\phi_{-\omega}^\text{II}(V)\right)^*\right\}\right],
    \label{eq:B2}
\end{align}
for the detector local mode given by
\begin{align}
	\hat{a}:=\int_0^\infty d\omega \left\{A(\omega)\,\hat{a}^\text{I}_\omega+B(\omega)\left(\hat{a}^\text{I}_\omega\right)^\dag+C(\omega)\,\hat{a}^\text{II}_\omega+D(\omega)\left(\hat{a}^\text{II}_\omega\right)^\dag\right\}.
\end{align}
This discussion is based on \cite{osawa2024final}.
The relation between canonical operators and creation and annihilation operators is given by
\begin{align}
	\hat{Q}=\frac{\hat{a}+\hat{a}^\dag}{\sqrt{2}},\quad \hat{P}=\frac{\hat{a}-\hat{a}^\dag}{\sqrt{2}i}.
	\label{eq:qpdet}
\end{align}
Whereas, by using profile functions, the canonical operators can be represented as
\begin{align}
    \label{eq:windetq}
	\hat{Q}&=\int_{-\infty}^\infty d{V} q(V)\int_0^\infty d\omega \left\{\pa_t\phi^\text{I}_\omega(V)\,\hat{a}^\text{I}_\omega+\pa_t\phi^\text{II}_{-\omega}(V)\,\hat{a}^\text{II}_{\omega}+\text{h.c.}\right\},\\
    \label{eq:windetp}
	\hat{P}&=\int_{-\infty}^\infty d{V} p(V)\int_0^\infty d\omega\left\{\pa_t\phi^\text{I}_\omega(V)\,\hat{a}^\text{I}_\omega+\pa_t\phi^\text{II}_{-\omega}(V)\,\hat{a}^\text{II}_\omega+\text{h.c.}\right\}.
\end{align} 
By comparing Eq.~(\ref{eq:qpdet}), Eq.~(\ref{eq:windetq}), and Eq.~(\ref{eq:windetp}),  profile functions $q_A(y), p_A(y)$ and weighting functions $A(\omega)$, $B(\omega)$, $C(\omega)$, and $D(\omega)$ are related as
\begin{align}
	\int d{V} q(V)\pa_t\phi^\text{I}_\omega(V)&=\frac{A(\omega)+B^*(\omega)}{\sqrt{2}},\quad
	\int d{V} p(V)\pa_t\phi_\omega^\text{I}(V)=\frac{A(\omega)-B^*(\omega)}{\sqrt{2}i},\\
    \int d{V}
    q(V)\pa_t\phi^\text{II}_\omega(V)&=\frac{C(\omega)+D^*(\omega)}{\sqrt{2}},\quad
	\int d{V} p(V)\pa_t\phi_\omega^\text{II}(V)=\frac{C(\omega)-D^*(\omega)}{\sqrt{2}i}.
\end{align}
The canonical commutation relation for the left-moving mode is
\begin{align}
	\left[\hat{\phi}(V),\hat{\Pi}(V')\right] 
	&=\int_0^\infty d\omega \left\{\phi^\text{I}_\omega(V)\left(\pa_t\phi^\text{I}_\omega(V')\right)^*-\pa_t\phi^\text{I}_\omega(V')\left(\phi^\text{I}_\omega(V)\right)^*\right\}\nn
	&\hspace{1cm}+\int_0^\infty d\omega \left\{\phi^{\text{II}}_{-\omega}(V)\left(\pa_t\phi^{\text{II}}_{-\omega}(V')\right)^*-\pa_t\phi^{\text{II}}_{-\omega}(V')\left(\phi^{\text{II}}_{-\omega}(V)\right)^*\right\}\\
 &\equiv\frac{i}{2}\delta\left(V-V'\right). \notag
\end{align}
Since the mode functions $\phi^\text{I}_\omega$ and $\phi_{-\omega}^{\text{II}}$ are restricted to  $V>0$ and $V<0$, respectively, the following normalization conditions hold:
\begin{align}
	\frac{1}{4}\delta(V-V')&=\Im\left[\int_0^\infty d\omega \,\phi^\text{I}_\omega(V)\left(\pa_t\phi^\text{I}_\omega(V')\right)^*\right]\quad (V,V'>0),\\
	\frac{1}{4}\delta(V-V')&=\Im\left[\int_0^\infty d\omega \,\phi^{\text{II}}_{-\omega}(V)\left(\pa_t\phi^{\text{II}}_{-\omega}(V')\right)^*\right]\quad (V,V'<0).
\end{align}
By using these identities, we obtain \eqref{eq:B1} and \eqref{eq:B2}.

\bibliography{mirror} 


\end{document}